\documentclass[preprint2]{aastex}

\shorttitle{Galaxy clustering with their properties}
\shortauthors{Kim et al.}

\begin{document}


\title{Linking galaxies to dark matter haloes at $z\sim1$ : dependence of
galaxy clustering on stellar mass and specific star formation rate}


\author{Jae-Woo Kim\altaffilmark{1}, Myungshin Im\altaffilmark{1}, Seong-Kook Lee\altaffilmark{1},
Alastair C. Edge\altaffilmark{2}, David A. Wake\altaffilmark{3,4}, 
Alexander I. Merson\altaffilmark{5},  and Yiseul Jeon\altaffilmark{1}}
\email{kjw0704@astro.snu.ac.kr, mim@astro.snu.ac.kr}

\altaffiltext{1}{Center for the Exploration of the Origin of the Universe, 
Department of Physics and Astronomy, Seoul National University, Seoul, Korea}
\altaffiltext{2}{Institute for Computational Cosmology, Department of Physics, 
University of Durham, South Road, Durham, UK}
\altaffiltext{3}{Department of Astronomy, University of Wisconsin, Madison, WI 53706, USA}
\altaffiltext{4}{Department of Physical Sciences, The Open University, Milton Keynes, UK}
\altaffiltext{5}{Department of Physics and Astronomy, University College London, Gower Place, London, UK}




\begin{abstract}
We study the dependence of angular two-point correlation functions on 
stellar mass ($M_{*}$) and specific star formation rate (sSFR) 
of $M_{*}>10^{10}M_{\odot}$ galaxies at $z\sim1$. 
The data from UKIDSS DXS and CFHTLS covering 8.2 deg$^{2}$ sample scales larger than 
100 $h^{-1}$Mpc at $z\sim1$, allowing us to investigate the correlation between clustering, $M_{*}$, and star formation through halo modeling. 
Based on halo occupation distributions (HODs) of $M_{*}$ threshold samples, we derive HODs for $M_{*}$ binned galaxies, and then calculate the $M_{*}/M_{\rm halo}$ ratio.
The ratio for central galaxies shows a peak at $M_{\rm halo}\sim10^{12}h^{-1}M_{\odot}$, and satellites 
predominantly contribute to the total stellar mass in cluster environments with $M_{*}/M_{\rm halo}$ values of 0.01--0.02.
Using star-forming galaxies split by sSFR, we find that 
main sequence galaxies ($\rm log\,sSFR/yr^{-1}\sim-9$) are mainly central galaxies in $\sim10^{12.5} h^{-1}M_{\odot}$ haloes 
with the lowest clustering amplitude, while lower sSFR galaxies consist of a mixture of 
both central and satellite galaxies where those with the lowest $M_{*}$ 
are predominantly satellites influenced by their environment. Considering the lowest $M_{\rm halo}$ samples in each $M_{*}$ bin, 
massive central galaxies reside in more massive haloes with lower sSFRs than low mass ones, indicating star-forming central galaxies evolve from a low $M_{*}$--high sSFR to 
a high $M_{*}$--low sSFR regime. We also find that the most rapidly star-forming galaxies 
($\rm log\,sSFR/yr^{-1}>-8.5$) are in more 
massive haloes than main sequence ones, possibly implying galaxy mergers in dense environments are driving the active star formation.
These results support the conclusion that the majority of star-forming galaxies follow secular 
evolution through the sustained but decreasing formation of stars.
\end{abstract}





\section{INTRODUCTION}

It is expected that small structures merge to form more massive ones in the lambda 
cold dark matter ($\Lambda$CDM) paradigm. Therefore small dark matter haloes 
are the seeds for larger structures. Galaxies form in these dark matter haloes through 
the binding of baryons and the cooling of gas \citep{whi78}. Since galaxies evolve in their host haloes, 
the distribution and evolution of galaxies are tightly related to their host dark matter haloes \citep{bau06}.
In the context of hierarchical structure formation models, massive dark matter haloes can contain 
many galaxies with a wide range in mass. Furthermore, the population of 
member galaxies depends on the properties of their host dark matter halo, because the potential well of the dark matter 
halo affects the properties of galaxies within it.

A popular method for measuring the distribution of galaxies is the two-point correlation 
function, which describes the excess probability of a galaxy pair over a random distribution on
specific scales \citep{pee80}. Recently, wide and deep surveys have provided opportunities to study the 
dependence of the clustering of galaxies on their various intrinsic properties such as color, luminosity, 
stellar mass and population \citep{nor01,nor02,zeh02,zeh05,coi08,ros09,loh10,ros10,zeh11}. As a result, 
it is known that redder, brighter or more massive galaxies are more strongly clustered than those 
having opposite properties.

For high redshift galaxies, one of the most efficient selection methods is to use their observed color.
Thus, many previous studies have applied various color cuts to select high redshift galaxies such as 
Extremely Red Objects \citep{els88,dad00,im02,roc02,roc03,yan04,bro05,kon06,gon09,kon09,kim11,pal13,kim14}, 
$BzK$ galaxies \citep{dad04,kon06,har08,mcc10,han12,mer13} and Distant Red Galaxies \citep{fra03,gra06,fou07,qua08,guo09}.
Their clustering properties also show similar trends to those of low redshift galaxies. Although 
color selection is efficient in isolating galaxies in a specific redshift range, a simple color 
cut often extracts a mix of galaxies with different properties and redshift.
Hence to correctly trace galaxy clustering
it is necessary to measure it from a sample containing galaxies with well defined 
intrinsic properties and a narrower range in redshift.

The halo occupation distribution (HOD) framework makes it possible to interpret the 
galaxy clustering in relation to their host dark matter haloes \citep{jin98,ben00,ma00,pea00,sel00,sco01,berl02,
coo02}. The HOD quantifies the mean number of central or satellite galaxies in a given halo mass \citep{krav04,zhe05}.
Based on the halo model analysis with multiwavelength datasets, many authors have reported that massive or luminous 
galaxies are found in more massive haloes \citep{zhe07,ros09,wak11,zeh11,cou12}. 

However, the HOD framework fits the number density of galaxies and their clustering simultaneously
so we require a large number of galaxies to sufficiently constrain the parameters of both. Also, 
to avoid the effects of cosmic variance, these galaxies need to be mapped over a large area of sky.
The lack of large, sensitive near-IR detectors has prevented identification of a 
large number of galaxies at $z\geq1$, where the bulk of stellar emission is observed in a near-IR regime. 
So far, previous work on stellar mass limited galaxies at high redshifts has been 
based on survey data with areas from a few hundreds arcmin$^{2}$ to $<$1.7 deg$^{2}$ \citep{men08,men09,fou10,har10,fur11,wak11,har13}. 
Furthermore, this small surveyed area makes it difficult to measure reliable clustering strengths on larger scales 
(a few tens $h^{-1}$Mpc), where the distribution of dark matter haloes is imprinted, since these surveys have covered at most 
$\sim$ 50 $h^{-1}$Mpc at $z\sim1$ on a side. Therefore it is important to perform
this analysis with homogeneous galaxy samples drawn from a wide-area near-IR survey.

In terms of galaxy evolution, different mechanisms play key roles at different epochs. \citet{pen10} 
proposed that mass quenching is important for all galaxies, but environment quenching dominates 
at low redshift and at lower masses. Also, there have been many results about the stellar mass function of 
passive galaxies from wide field optical--near-IR datasets, showing relatively mild evolution of the most
massive galaxies but a dramatic change for low mass ones 
\citep{dro09,ilb10,bez12,ilb13,mou13,muz13,tin13,tom14}, which is consistent with the model in \citet{pen10}. In addition, the evolution of 
the luminosity function of Luminous Red Galaxies (LRGs) follows a passive evolution model at $z<0.6$ 
\citep{wak06}. \citet{sco13} reported that the fraction of early type galaxies increases from 30\% at $z\sim1.1$ 
to 80\% at $z\sim0.2$ in the densest regions, but from 30\% to only 50\% in low density regions. 
These results suggest that the dependence of galaxy properties on their host haloes at $z\sim1$ is different from that 
in the local universe.

Furthermore, the relation between galaxy properties and their environment at $z\sim1$ is still controversial. 
\citet{elb07} and \citet{coo08} found a reversed relation between star formation rate (SFR) and environment at $z\sim1$, 
meaning a higher SFR was observed in the highest density regions. \citet{sco13} found the evolution of the relation as a function of redshift, 
and recover a weak or no dependence of SFR on environment at $z\sim1$. On the other hand, \citet{coo10}, \citet{chu11} and 
\citet{qua12} reported that the color (or SFR)--density relation persists out to $z\sim1.5$.
\citet{tin13} argued that the central 
galaxies in low mass haloes are likely to be star-forming galaxies at $z\sim1$, and their evolution contributes to the observed change in the red 
sequence. However, there are not many results connecting galaxy properties with their host haloes at $z\sim1$, 
especially detailed, statistical studies based on a large sample from wide area surveys ($\gtrsim$2--3deg$^{2}$). \citet{mos13} measured the 
clustering strength of galaxies with various criteria such as stellar mass, SFR and specific star formation rate (sSFR) at 
$z\sim1$ from a small area spectroscopic survey. We re-address their work with a much larger photometric dataset making it 
possible to split galaxies into finer sub-samples. 

In this work, we use wide and deep multiwavelength datasets with $ugrizJ$- and $K$-bands based on UK Infrared Telescope (UKIRT) 
Infrared Deep Sky Survey (UKIDSS) Deep eXtragalactic Survey (DXS) 
and Canada-France-Hawaii Telescope (CFHT) Legacy Survey (CFHTLS)--Wide. The catalog covers 8.2 deg$^{2}$ with the limit magnitude of $J_{AB}=23.2$. This is one 
of the best datasets to investigate the clustering properties of homogeneous galaxy sub-samples and to minimize 
the influence of cosmic variance, thanks to its unique combination of depth and area. Furthermore, these data 
allow us to link galaxies with various criteria to their 
host haloes separately. Using this catalog, we measure the angular 
two-point correlation function of $z\sim1$ galaxies split into several sub-samples based on stellar mass and 
sSFR. Additionally, we fit a halo model and measure the bias factor with the 
measured correlation function in order to link galaxies with their host dark matter haloes.

In \S~\ref{data}, we briefly describe each survey and how the catalog was generated. In 
\S~\ref{method} we note the methods applied to select samples, to measure clustering and to model 
the HOD. The dependence of galaxy clustering on stellar mass and the stellar mass to halo mass relation 
are described in \S~\ref{clmass}. We present the dependence on sSFRs in \S~\ref{clssfr}. We also relate these 
results to the evolution of galaxies in \S~\ref{discuss}, and finally summarize this work in \S~\ref{sumcon}.
Throughout this paper, $M_{*}$ indicates a stellar mass of galaxies and $M_{\rm halo}$ means a dark matter halo 
mass. We assume a flat $\Lambda$CDM cosmology: $\Omega_{m}=0.27$, $\sigma_{8}=0.8$, 
$H_{0}=100 h$ km s$^{-1}$ Mpc$^{-1}$ with $h=0.71$. The photometry is quoted in the AB system.

\section{DATA}\label{data}


\subsection{{\it UKIDSS DXS}}\label{dxs}

The DXS (A. C. Edge et al. 2015, in preparation) is a sub-survey 
of the UKIDSS which was performed from 2005 to 2012
\citep{law07} using the UKIRT. The DXS images were obtained 
using the Wide Field Camera (WFCAM, Casali et al. 2007) composed of four Rockwell 
Hawaii-II 2K$\times$2K array detectors covering four 13.7$\times$13.7 arcmin$^{2}$ 
regions. Since WFCAM has a relatively large pixel scale as 0\arcsec.4/pixel, 
a microstepping technique has been applied so that a science image has 0\arcsec.2/pixel 
and avoids an undersampled point spread function.

The DXS maps $\sim$35 deg$^{2}$ composed of 4 different 
8.75 deg$^{2}$ patches (XMM-LSS, Elain-N1, Lockman Hole and SA22) with aimed
depths of $J_{AB}=23.2$ and $K_{AB}=22.7$ at a 5$\sigma$ point-source sensitivity.
The actual data show a 90\% point-source completeness at these magnitudes \citep{kim11}. 
The scientific goals of the survey are to determine the abundance of galaxy clusters at $z>1$,
to understand the clustering of galaxies, and to investigate the census of the luminosity
density in star formation.

In this study, we deal with the SA22 field centered on $\alpha=$22$^{h}$17$^{m}$00$^{s}$
and $\delta=+$00\arcdeg20\arcmin00\arcsec (J2000). In the whole surveyed area (3.4$\times$2.6 deg$^{2}$), we perform 
our analysis with images from the UKIDSS data release 9 (DR9) covering $\sim8$ deg${^2}$, and one remaining WFCAM field 
($\sim$0.7 deg$^{2}$) which was not released in DR9 comes from DR10. The area coverage corresponds to roughly 
140 $h^{-1}$Mpc $\times$ 107 $h^{-1}$Mpc at $z=1$. The average seeing is $\sim0\farcs8$ 
in both $J$ and $K$. The photometric and astrometric solutions are based on the output 
from the standard pipeline, and are accurate to better than 2\% and $\sim$0\arcsec.05 \citep{dye06,law07}.

\subsection{{\it CFHTLS}}\label{cfht}

The CFHTLS\footnote{http://www.cfht.hawaii.edu/Science/CFHTLS/} is 
a set of deep and wide optical surveys performed using 
the MegaCam camera mounted on the CFHT with $ugriz$ 
filters. Of the three surveys that constitute the CFHTLS, we deal with the CFHTLS--Wide W4 field which 
covers the DXS SA22 area. The CFHTLS W4 field covers 25 deg$^{2}$ with limiting magnitudes 
(50\% completeness for point sources) of 
$u\sim26.0$, $g\sim26.5$, $r\sim25.9$, $i\sim25.7$ and $z\sim24.6$ \citep{gwy12}.
For this work, we use CFHTLS images taken from the MegaPipe data pipeline at the Canadian 
Astronomy Data Centre \citep{gwy12}. The image reduction procedure, as well as photometric and astrometric 
calibrations are well described in \citet{gwy12}.

The UKIDSS DXS area is located in the south-east corner of the CFHTLS W4 field. Thus we extract 
only the sub-region that overlaps with the UKIDSS DXS field. We use the images in all the CFHT filters, 
i.e., $u, g, r, i,$ and $z$.

\subsection{{\it Catalog}}\label{cat}

Our main goals in this work are to measure the clustering of galaxies at $z\sim1$ and to 
investigate how the clustering of galaxies correlates with the star formation activity 
and stellar mass of these galaxies.
Therefore it is important to accurately determine colors of galaxies to perform spectral energy 
distribution (SED) fits to estimate galaxy properties as well as to determine their photometric redshifts. 
In an attempt to improve the photometric accuracy, we generate a new catalog instead of using the released catalogs 
from the UKIDSS team via WFCAM Science Archive\footnote{http://surveys.roe.ac.uk/wsa/}. 
The new catalog of objects was constructed using the procedure below.

First, since fluxes must be measured from the same region of galaxies at different bands to obtain 
accurate color, images are convolved through a Gaussian filtering to unify the FWHM. 
The worst seeing condition of our UKIRT dataset is used as a reference, which 
corresponds to a FWHM of 1\farcs1. Then we resample all images to have the WFCAM field of view with 
a four pointing observation and CFHT pixel scale, using the Swarp software \citep{ber02}. 
Hence, we have 12 images for each band, each covering 0.8 deg$^{2}$. We refer to each 0.8 deg$^{2}$ image 
as a tile. The coordinate offsets ($\Delta$R.A. and $\Delta$decl.) between both surveys are 0\farcs05 and 
0\farcs04, respectively, which are accurate enough to align images.

Second, we run the SExtractor software \citep{ber96} in dual mode on each tile of the covered area. 
The unconvolved $J$-band image is used for the detection and the measurement of 
AUTO magnitudes which we consider a proxy of total magnitudes of an object. In order to measure the color of 
each object, 2\arcsec\,diameter apertures are used to derive aperture magnitudes on the Gaussian filtered images 
in all bands.

Third, spurious objects such as cross-talk and objects on diffraction spikes are removed. 
It is well known that bright stars make cross-talk at specific positions on the WFCAM chip \ \citep{dye06}. 
In addition, the unreliable regions of haloes or diffraction spikes of bright stars are 
also masked. We follow the algorithm in \citet{kim11} for reducing spurious objects. The 
masking region for CFHTLS is taken from the Megapipe data pipeline.

\begin{figure}[h]
\centering
\includegraphics[scale=0.35]{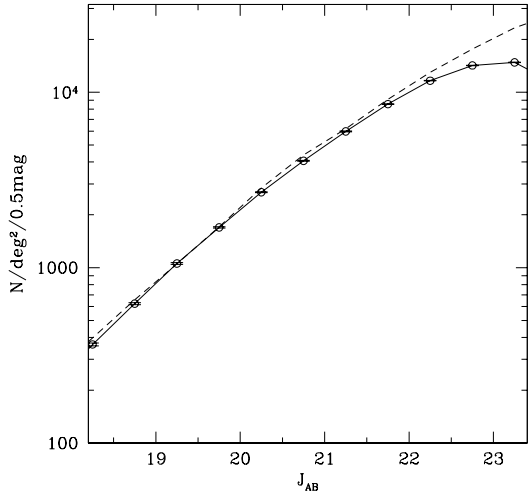}
\caption{Number counts of galaxies in the DXS SA22 field. The solid line is 
for all galaxies in this work. Dashed line is the result from \citet{jar13}. 
Our result is consistent with previous results up to $J_{AB}\sim22.5$. 
\label{fignd}}
\end{figure}

\begin{figure}[h]
\includegraphics[scale=0.35]{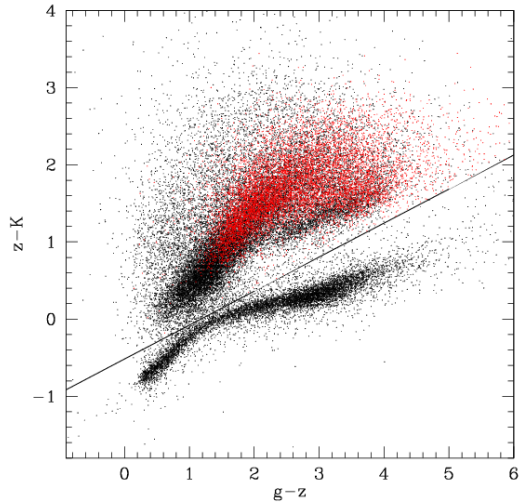}
\caption{The ($z-K$) versus ($g-z$) two-color diagram. We extract objects in a single tile 
(0.8 deg$^{2}$) with $J<22.5$ for a display purpose (black dots). The line indicates the criterion 
to distinguish galaxies from stars. Stars fall below the solid line. Red dots are objects satisfying our selection criteria for the analysis in 
section 3.1. For the display purpose, $\sim$10,000 objects are shown. \label{figgzk}}
\end{figure}

Finally, all sub-catalogs from different tiles are merged to create a single, master catalog. If objects have been detected in an overlapping 
region of different tiles, a 1\arcsec\,matching radius was applied to identify objects in common for both
tiles and the weighted-mean of 
fluxes from different tiles are assigned as the final flux of uniquely identified
objects. Additionally, the Galactic extinction is corrected based 
on the dust map from \citet{sch98}. 
In total, $\sim$0.86 
million objects are found in the 8.2 deg$^{2}$ UKIDSS DXS area. Figure~\ref{fignd} shows the number counts of 
all galaxies in the DXS area (solid line). Stars were excluded using ($g-z$) and 
($z-K$) colors, which is similar to the $BzK$ diagram \citep{dad04,oi14}. Figure~\ref{figgzk} displays all 
detected objects in a single tile (0.8 deg$^{2}$) with $J<22.5$ as an illustration. The line indicates 
the criterion, $(z-K)=0.44(g-z)-0.52$, which we apply to distinguish galaxies from stars. 
The number counts from \citet{jar13} are also indicated with a dashed line in Figure~\ref{fignd}. 
Our result shows a relatively good agreement over the whole magnitude range. 
However, we may miss some galaxies in the faint regime ($J_{AB}>22.5$) due to the 
relatively shallow depth of CFHTLS and the associated incompleteness. 
In fact, the distribution of objects shows a large 
scatter in their colors in the faintest magnitudes, where the magnitude of optical 
bands is close to the limiting magnitude, making it difficult to distinguish galaxies 
from the stellar locus. In addition, the 90\% completeness for extended sources is $\sim$0.3-0.5 magnitude brighter than that for point sources. 
We note that we will apply additional criteria of galaxy stellar mass ($M_{*}>10^{10}M_{\odot}$) and photometric redshift ($0.8<z<1.2$) 
for our analysis, so that we select relatively bright galaxies which are well separated from the stellar locus
so we expect negligible stellar contamination in our sample.
Red dots in Figure~\ref{figgzk} are the objects satisfying our selection criteria. For the display purpose, we display 
a portion of full samples.

\section{METHODS}\label{method}

\subsection{{\it Selection}}

This section describes the basic measurement for galaxy properties and 
the criteria for our sample selection with the multiwavelength catalog from $u$-band 
to $K$-band corresponding to the rest-frame UV to near-IR for $z\sim1$ galaxies.

\subsubsection{{\it Photometric Redshift}} \label{bozomath}

The redshift information is crucial to estimate galaxy properties and 
to investigate the clustering of galaxies in a specific redshift range. Using our
multiwavelength catalog, we 
estimate the photometric redshift of each object.

\begin{figure}[h]
\includegraphics[scale=0.35]{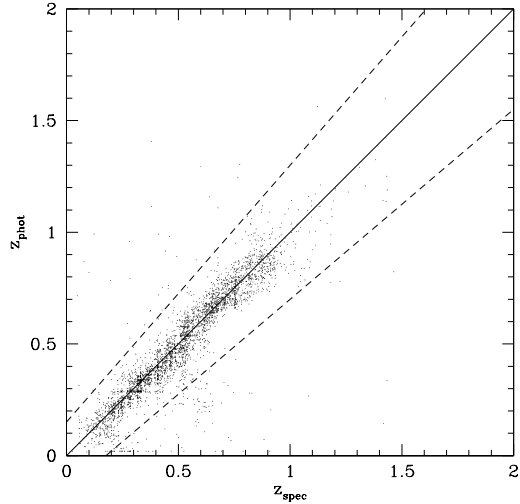}
\caption{Photometric redshifts ($z_{phot}$) versus spectroscopic redshifts ($z_{spec}$) for 
objects observed by the VVDS survey. The dashed lines indicate $\Delta z/(1+z)=0.15$.
\label{figszpz}}
\end{figure}

We used the {\it Le Phare} \citep{arn99,ilb06} software to derive photometric redshifts. 
We used 66 SED templates applied for the CFHTLS--Deep fields in \citet{ilb06}. The templates are based on 
Ell, Sbc, Scd and Irr spectra from \citet{col80} and a starburst SED from \citet{kin96}, and cover the 
wavelength range from rest-frame UV to near-IR (see Ilbert et al. 2006 for more details). The 
{\it Le Phare} code produces offsets in magnitude in each band after running the code on a training set 
of galaxies for which spectroscopic redshifts are available. The application of the magnitude offset improves 
the photometric redshift accuracy, and as a training set, we used
the VIMOS-VLT Deep Survey (VVDS) wide which is a spectroscopic survey mapping 4 deg$^{2}$ of the 
UKIDSS DXS SA22 area for $I<22.5$ objects \citep{lef05,gar08}. We select 3609 galaxies in the SA22 field 
having reliable spectroscopic redshift information as a training set. First, we ran the 
{\it Le Phare} software for the cross-matched objects between UKIDSS DXS 
and VVDS to calculate the magnitude offset for each band compared to those from templates, which are $\sim$0.25 mag for $J$-band 
and less than 0.08 mag for the others. The large offset in $J$-band appear in other works (e.g., \citet{ilb09}), and 
can be understood as due to a template mismatch. In this case, the redshift was fixed with that from VVDS. 
After finding and then applying these magnitude offsets to the objects in the catalog, the {\it Le Phare} software 
was run again for all detected objects. Figure~\ref{figszpz} displays 
the comparison between measured photometric redshifts ($z_{phot}$) and VVDS spectroscopic 
redshifts ($z_{spec}$) for the cross-matched galaxies. We find that the normalized median absolute deviation 
of photometric redshift in $\Delta z/(1+z)$ is $\sim$0.038. For galaxies with $0.8<z_{spec}<1.2$ that are 
studied in this work, the uncertainty is $\sim$0.042. The fraction of 
outliers ($|\Delta z/(1+z)|>0.15$) is $<5$\% for both cases. 
The dashed lines in Figure~\ref{figszpz} show where $\Delta z/(1+z)=0.15$. 
Hereafter the term redshift ($z$) indicates the photometric result ($z_{phot}$). We use photometric redshifts only 
for the analysis. We also note that the magnitude offset is applied for the estimation of photometric redshifts only. 
The magnitude offset improves the photometric redshift estimation by removing a systematic offset of $\Delta z=$0.04.

Finally, we apply a redshift cut of $0.8<z<1.2$, and remove Galactic stars based on the $gzK$ diagram as described 
in \S~\ref{cat}. Note that no magnitude cuts are applied when deriving photometric redshifts. 
In total 141,947 galaxies in this redshift range are used for estimating stellar masses and 
star formation activity. 

In order to check how efficiently the use of photometric redshifts captures galaxies at $0.8<z<1.2$,
we use a galaxy mock catalog from the {\tt GALFORM} semi-analytical model \citep{col00,mer13}.
For this test, $J$-band magnitudes are randomly scattered in the photometric uncertainty ranges as a function of 
magnitude, and we 
selected galaxies satisfying $J<23.2$, which is the magnitude limit in the observed catalog.
We also randomly assign the redshift uncertainty to the mock galaxies. Then we compare the number of galaxies selected with modified 
redshifts to that with true redshifts. We find through this test that the use of photometric redshifts can recover 90\% of 
galaxies with true redshifts at $0.8<z<1.2$, while the interlopers (foreground or background galaxies) are about 13\% among 
galaxies at $0.8<z<1.2$.

\subsubsection{{\it SED Fit}}\label{sedfit}

\begin{figure}[h]
\centering
\includegraphics[scale=0.4]{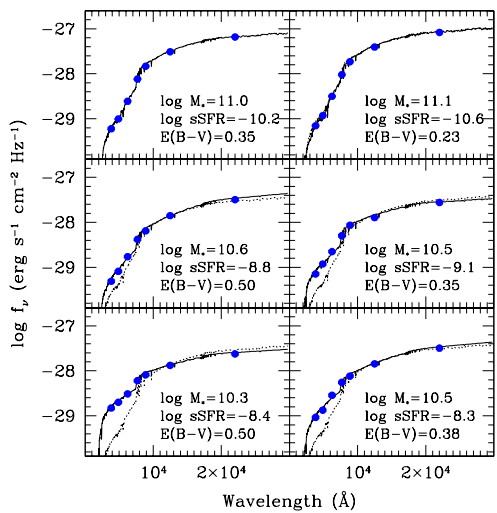}
\caption{Examples of best fit SEDs (solid line) with observed fluxes (points) for galaxies at $z\sim1$. 
Stellar mass and sSFR values of each galaxy are noted in each panel with units in $M_{\odot}$ and $yr^{-1}$, respectively. 
The dotted line is the passive galaxy SED in the top-left panel with normalizing to the flux at 1.2 $\mu$m of each object.
\label{figsedexam}}
\end{figure}

In order to estimate stellar masses ($M_{*}$) and SFRs of galaxies, we fit 
model templates of synthetic stellar populations to the multi-band photometry following the
algorithm of \citet{lee10,lee14}. Here we briefly note the assumptions made in this analysis.
We use SED templates from \citet{bru03} with a \citet{cha03} initial mass function.
We assume a delayed star formation history with an age ($t$) from 0.1 Gyr to the age of Universe 
at the redshift of the galaxy in question and a star formation timescale parameter ($\tau$) from 
0.1 to 10 Gyr. 
Also the \citet{cal00} dust attenuation curve is assumed for internal extinction.
The reddening parameter of $E(B-V)$ ranges from 0.0 to 1.5 with a step size of 0.025.
The metallicity was allowed to have values of 0.2, 0.4, 1.0 and 2.5 $Z_{\odot}$.
The SED fit returns the best fit parameters such as $M_{*}$, SFR, age, $\tau$ and 
$E(B-V)$. The SFR is defined as the averaged one over recent 100 Myr, based on the reasoning of 
\citet{lee09b}.
Figure~\ref{figsedexam} displays examples of SED fits for galaxies at $z\sim1$. The solid line is the best fit SED, and points are observed 
fluxes at each band. The dotted line shows the SED in the top-left panel. For comparison purposes, 
we normalize the SEDs to the flux at observed frame 1.2 $\mu$m of each object.

\begin{figure*}
\centering
\includegraphics[scale=0.4]{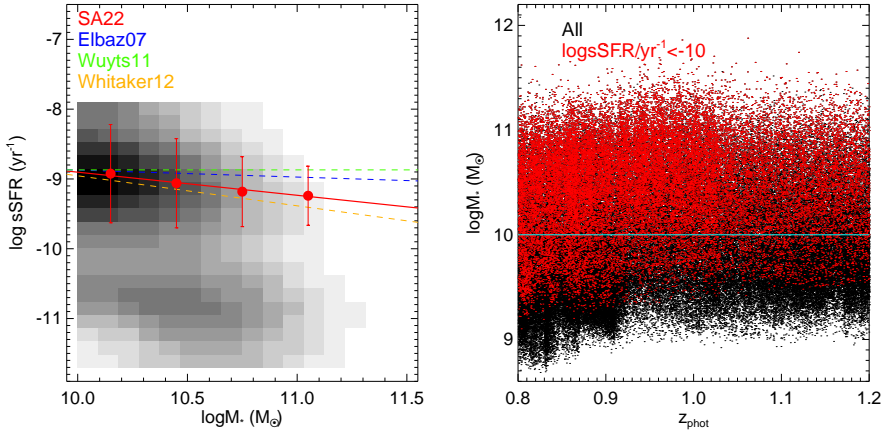}
\caption{
{\bf Left:} the sSFR-stellar mass distribution of galaxies at $0.8<z<1.2$ in the SA22 field. The 
red line indicates the sSFR of the main sequence galaxies in this work. Red points present the peak location for star-forming galaxies 
derived by a double Gaussian fit to galaxies in different stellar mass bins (see the text for more details). The blue, green and orange lines
are results from \citet{elb07}, \citet{wuy11} and \citet{whi12}, respectively.
{\bf Right:} the stellar masses of galaxies in $0.8<z_{phot}<1.2$. Black 
and red dots are all and $\rm log\,sSFR/yr^{-1}<-10$ objects, respectively. The cyan line is the stellar mass 
cut ($M_{*}=10^{10}M_{\odot}$) applied in this work. For the display purpose, 
objects in 1 deg$^{2}$ of surveyed area are plotted.\label{figmssfr}}
\end{figure*}

The left panel in Figure~\ref{figmssfr} displays sSFR versus $M_{*}$ of galaxies at 
$0.8<z<1.2$ in the SA22 field. We find that the fit of a power-law ($\rm log\,sSFR=\alpha \rm log\,M_{*}+\beta$) to
the main sequence of star-forming galaxies has a slope of
$\alpha=-0.33\pm0.03$ and an intercept of  $\beta=-5.59\pm0.31$ (red line). For the estimation, we perform a double Gaussian 
fit to the sSFR distribution of galaxies in 5 different stellar mass bins from $M_{*}=10^{10}M_{\odot}$ to $M_{*}=10^{11.3}M_{\odot}$ with 
a width of 0.3 dex. Then the power-law is fit to the peak location of sSFRs for star-forming galaxies only.
As a reference, 
the main sequence of star-forming galaxies at $0.8<z<1.2$ with the power-law slope of -0.10 from \citet{elb07} is displayed with a blue line. 
In addition, we also plot the main sequence of star-forming galaxies at $0.5<z<1.5$ 
taken from \citet{wuy11} (green line), although the power-law slope of unity for the 
SFR-$M_{*}$ relation was assumed in their analysis. 
On the other hand, \citet{whi12} suggested the significant evolution in the power-law 
slope of the SFR-$M_{*}$ relation with redshifts. The relation at $z=1$ derived by equations in \citet{whi12} is shown as an orange line with 
the slope of -0.43, which is similar to our estimate. We also note that our relation is similar to that in \citet{whi12}, and 
our sample includes more high mass galaxies than \citet{whi12}.
This figure shows two loci of galaxies, one with $\rm log\,sSFR/yr^{-1}>-10$, and another with $\rm log\,sSFR/yr^{-1}<-10$. 

Based on $J_{AB}<23.2$ galaxies at $1.15<z<1.2$, the 80\% percentile of the stellar mass distribution is 
$M_{*}\sim10^{9.5}M_{\odot}$. However, we may miss a fraction of the lower mass, passive 
galaxies due to the relatively shallow optical dataset. Therefore we use only galaxies more 
massive than $M_{*}=10^{10}M_{\odot}$ for our analysis. 
The right panel in Figure~\ref{figmssfr} shows the stellar mass of detected galaxies as a 
function of redshift with $\rm log\,sSFR/yr^{-1}<-10$ galaxies plotted in red and all other galaxies in black.
The horizontal 
line is the stellar mass cut of $M_{*}=10^{10}M_{\odot}$. Consequently, we extract 66,864 galaxies 
in $0.8<z<1.2$ with $M_{*}>10^{10}M_{\odot}$ for this work.

\subsection{{\it Angular Two-point Correlation Function}}\label{atcf}

One of the simplest ways to measure the clustering of galaxies is the two-point correlation 
function, which is the excess probability of finding a galaxy pair over a random distribution at a given 
scale \citep{pee80}. Here, we measure the angular two-point correlation function of galaxies at 
$0.8<z<1.2$, using the estimator introduced by \citet{lan93} : 
\begin{equation}
w_{\rm obs}(\theta)=\frac{DD(\theta)-2DR(\theta)+RR(\theta)}{RR(\theta)},
\end{equation}
where DD is the number of galaxy pairs in $\theta\pm\Delta\theta$ in the observed data. In this work, the bin width is 
chosen as $\Delta \log\theta/ \rm degree=0.15$. DR and RR are the number of galaxy-random and random-random pairs, 
respectively. We generated 30 different random point catalogs having the same areal coverage and angular mask as observed with 
each random point catalog containing a similar number of random points to that 
of observed galaxies. The errors on the  two-point correlation functions and the covariance matrices were estimated 
by the Jackknife resampling method, after splitting the UKIDSS DXS area into 48 sub-fields. 

Although our data cover a wide area, it is hard to avoid the effect caused by the finite 
survey area, which is referred to as the integral constraint (IC, Groth \& Peebles 1977) which is additive, i.e., 
$w(\theta)=w_{\rm obs}(\theta)+\rm IC$. This effect can be corrected by the empirical method in \citet{roc99} with an
assumption of the actual shape of two-point correlation function. 
We apply 
two different approaches for stellar mass limited samples in \S~\ref{clmass} and sSFR binned samples in \S~\ref{clssfr}.
For stellar mass limited samples, 
we use each two-point correlation function obtained from the halo model with observed parameters
and then calculate the integral constraint with the empirical equation in \citet{roc99}. This process is included in the model fitting procedure 
to observed correlation functions, and taken into account to find the best fit HOD parameters (see Wake et al. 2011). 
However, the integral constraint must be measured differently from stellar mass limited samples for sSFR binned samples, since we do 
not perform the halo modeling for this selection. As introduced in 
\citet{kim11}, we assume the empirical functional form as the true correlation function, and then use the 
iterative method in \citet{roc99}. In this work, we assume the functional form of 
$w({\theta})=\alpha_{1}\theta^{-\beta_{1}}+\alpha_{2}\exp(-\beta_{2}\theta)$ \citep{kim14}. 
The combination of a power-law and an exponential component is necessary to describe small scale clustering of galaxies (power-law component) and large scale 
halo-to-halo clustering (exponential component), simultaneously. We note that the integral 
constraints of sub-samples range from 0.001 to 0.013 and that a more clustered sample tends to have a larger integral 
constraint value. Furthermore, for the stellar mass limited samples, the IC values measured by using the halo model and the functional form 
show the differences of $<$0.0017.

\subsection{{\it HOD Fit}}\label{hod}

Since galaxies reside in dark matter haloes, the distribution of dark matter haloes or density peaks are imprinted on 
the clustering of galaxies. In this context, we are able to link galaxies with dark matter 
haloes through the halo model (see Cooray \& Sheth 2002 for a review). Here, we apply a halo 
model using the Halo Occupation Distribution (HOD) to study the relation between galaxies 
and their host dark matter haloes. This model has been widely applied to various galaxy 
populations \citep{zhe07,bla08,wak08,ros09,saw11,wak11,zeh11,kra13,niko13,kim14}. 

In order to model the best fit correlation function, we have to parameterize the basic components of the 
halo model such as the number density of haloes, the satellite distribution in haloes, the halo 
bias and the mean number of galaxies at a given halo mass. First, we adopt the halo mass function 
($n(M_{\rm halo})$) and the halo bias function ($B(M_{\rm halo})$) from \citet{tin10} for the number 
density and bias of haloes, respectively. 
Second, the distribution of satellites is assumed to follow the NFW profile \citep{nav97} with the concentration parameter 
depending on redshift \citep{bul01,zeh04,bla08,ros09}. 
Since $n(M_{\rm halo})$, $B(M_{\rm halo})$ and the NFW profile depend on redshift, we determine these at the mean redshift of our 
sample, i.e., $z=1$. 
Finally, we specify the mean number of galaxies at a given halo mass ($N(M_{\rm halo})$), which in turn is parameterized 
for central galaxies ($N_{c}$) and satellites ($N_{s}$) separately and was introduced by \citet{zhe05}, as 
\begin{equation}
N(M_{\rm halo}) = N_c(M_{\rm halo}) + N_s(M_{\rm halo}),
\end{equation}
with 
\small
\begin{equation}
N_c(M_{\rm halo}) = 0.5 \left[ 1 + {\rm erf}\left(\frac{{\log_{10}} (M_{\rm halo}/M_{\rm cut})}{\sigma_{\rm cut}}\right)\right]
\end{equation}
\normalsize
and
\footnotesize
\begin{equation}
N_s(M_{\rm halo}) = 0.5 \left[ 1 + {\rm erf}\left(\frac{{\log_{10}} (M_{\rm halo}/M_{\rm cut})}{\sigma_{\rm cut}}\right)\right] \left(\frac{M_{\rm halo}-M_1}{M_0}\right)^{\alpha}, 
\end{equation}
\normalsize
where $M_{\rm halo}$ is a dark matter halo mass. $M_{cut}$ and $\sigma_{cut}$ define the transition 
halo mass and shape of HODs for central galaxies. $M_{1}$ is the truncation mass for satellites, and $M_{0}$ and $\alpha$ are the threshold halo mass and the slope 
for HODs of satellites, respectively (see also \citet{zhe05} and \citet{wak11}).

Our survey area of $\sim$140$h^{-1}$Mpc at $z=1$ on one side is not wide enough to fully constrain all five
free parameters ($M_{cut}$, $\sigma_{cut}$, $M_{0}$, $M_{1}$ and $\alpha$) simultaneously.
In previous 
work \citep{zhe07,bro08,zeh11}, it was reported that $M_{1}$ is poorly constrained but similar 
to $M_{cut}$ based on the SDSS data. Therefore we follow the relation, ($M_{1}=M_{cut}$), in this work.
In addition, $M_{cut}$ can be determined by matching the number density of galaxies with 
given parameters as applied in \citet{ros09}. Consequently, we have just three 
free parameters ($\sigma_{cut}$, $M_{0}$ and $\alpha$) to model the real-space correlation 
function. 

We assume that the mean number of central galaxies is unity beyond a specific halo mass. 
Recent work based on cosmological simulations has argued that this can be below unity even 
in very massive haloes due to active galactic nucleus (AGN) feedback \citep{kimhs09, gon11, con13}, if the stellar mass threshold is chosen to be very large. 
However, the effect of AGN feedback is still controversial 
and there is a debate as to whether this is positive or negative on short timescales ($<100$~Myr). 
Therefore we do not 
consider this effect in this work. Furthermore, \citet{zen13} pointed out that any halo model that does not take into account 
the assembly bias leads to a systematic error on the fitted result. Although this effect is significant for red populations, 
it is much reduced when considering all galaxies. 

Using the basic components mentioned above, we follow the scheme of \citet{ros09}. Briefly, 
we model power spectra contributed by galaxies in the same halo (1-halo term) and in different 
haloes (2-halo term). The 1-halo term is distinguished into central-satellite and 
satellite-satellite pairs again. In order to consider the underlying dark matter, the matter power 
spectrum is generated by the `CAMB' software package \citep{lew00} including the formulae of \citet{smi03} to 
model nonlinear growth. Then the modeled power spectrum is transformed to the real-space 
correlation function using the Limber equation \citep{lim54} to project the modeled 
real-space correlation function to angular space. The redshift distribution of each sub-sample is generated not by the 
best fit photometric redshift, but by the possible redshift distribution of each galaxy. We adopt 90\% redshift ranges from the 
{\it Le Phare} software for each galaxy, then assume the Gaussian distribution above and below the redshift separately. This distribution of each 
galaxy is stacked to produce the redshift distribution. 
Finally, the modeled angular correlation function 
is fitted to the observed correlation function with the covariance matrix derived in \S~\ref{atcf} to find the best fit parameters. We fit the 
modeled correlation function to the observed one over the angular range $3\farcs2<\theta<0\arcdeg.7$, where the influence of the integral constraint is minimized.

Additionally, the effective halo mass ($M_{eff}$), the effective bias ($b_{g}$) and the fraction 
of central galaxies ($f_{cen}$) are derived from the best fit parameters with 
\small
\begin{equation}
M_{\rm eff} = \int {\rm d}M_{\rm halo} M_{\rm halo} n(M_{\rm halo})N(M_{\rm halo})/n_g,
\end{equation}
\begin{equation}
b_{\rm g} = \int {\rm d}M_{\rm halo} B(M_{\rm halo}) n(M_{\rm halo})N(M_{\rm halo})/n_g
\end{equation}
\normalsize
and
\small
\begin{equation}
f_{\rm cen} = \int {\rm d}M_{\rm halo} n(M_{\rm halo}) N_c(M_{\rm halo})/n_g,
\end{equation}
\normalsize
where $n_{g}$ is the number density of galaxies, which is fixed to the observed value in this work.

We perform the halo modeling for only galaxies selected above a stellar mass threshold, 
since this model is appropriate for mass or luminosity limited samples. For the galaxies 
in stellar mass bins, the difference of HODs between two mass thresholds 
is calculated (see \S~\ref{mbin}).

\begin{figure*}
\centering
\includegraphics[width=100mm]{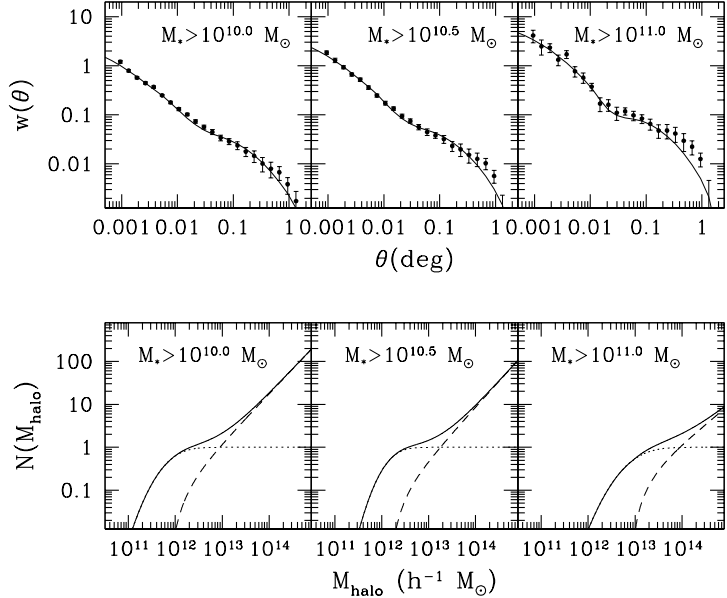}
\caption{The upper panels show angular two-point correlation functions of galaxies 
selected by stellar mass thresholds indicated in each panel (points with error bars). 
The solid line is the best fit halo model. The lower panels present the best fit 
HODs (solid line). Dotted and dashed lines indicate HODs for central galaxies and 
satellites, respectively. It is clear that more massive galaxies are in more massive haloes on 
average.\label{figmth}}
\end{figure*}

\section{CLUSTERING WITH STELLAR MASS}\label{clmass}

In this section, we describe the results from fitting the halo model to galaxies split by different stellar mass 
thresholds of $M_{*}>10^{10.0}$, $10^{10.5}$ and $10^{11.0}M_{\odot}$ (\S~\ref{mth}). Then the HODs for 
galaxies in different stellar mass bins ($10^{10.0}M_{\odot}<M_{*}<10^{10.5}M_{\odot}$, 
$10^{10.5}M_{\odot}<M_{*}<10^{11.0}M_{\odot}$ 
and $M_{*}>10^{11.0}M_{\odot}$) are compared in \S~\ref{mbin}. Finally based on these HODs, the ratio 
between stellar mass and halo mass is discussed in \S~\ref{mratio}

\subsection{{\it Mass Threshold}}\label{mth}

In order to investigate the dependence of the HOD parameters on stellar mass, we split the selected galaxies at
stellar mass thresholds of $M_{*}>10^{10}$, $10^{10.5}$ and $10^{11}M_{\odot}$. In total, 66,864, 
29,250 and 4,564 galaxies are selected for each stellar mass threshold, respectively. The upper panels of 
Figure~\ref{figmth} show the measured angular two-point correlation function for each sub-sample 
(points with error bars). The relatively large error for $M>10^{11.0}M_{\odot}$ comes from the 
significantly smaller number of galaxies. 
The solid line is the best fit halo model. As seen in Figure~\ref{figmth}, 
the halo model reproduces the observed angular correlation function well. It also shows 
a relatively good fit even beyond the fitted range ($\theta>0^{\circ}.7$)
and a more pronounced break between the 1- and 2-halo components as the 
balance of central to satellite galaxies changes with stellar mass.
The best fit parameters 
and derived quantities are listed in Table~\ref{tparam}.

The lower panels of Figure~\ref{figmth} display the best fit HODs for each sub-sample (solid line). 
The dotted and the dashed lines are HODs for central galaxies and satellites, respectively. It is 
clear that HODs for lower stellar mass galaxies extend down to a lower halo 
mass regime. This is well described by the best fit parameters of two halo mass thresholds ($M_{cut}$ and $M_{0}$), 
which decrease for low mass galaxies.
In addition, the HOD of more massive central galaxies shows a gentler transition shape compared to
less massive central galaxies as already reported by \citet{zeh11}. The best fit parameter $\sigma_{cut}$ describing 
the transition shape is 0.5 for $M_{*}>10^{10}M_{\odot}$ and 0.6 for $M_{*}>10^{11}M_{\odot}$. In order to check 
the effect of the galaxy number density to the best fit parameters, we perform the fit with four free parameters 
($\sigma_{cut}$, $M_{0}$, $\alpha$ and $M_{1}$). In this case, $n_{g}$ does not directly constrain $M_{1}$, but is calculated by 
the best fit parameters for matching the clustering only. Although, the parameters show different values from 3 parameter fits, 
the trend is exactly same. These results are presented in Appendix~\ref{app1}.

\begin{figure}[h]
\includegraphics[scale=0.7]{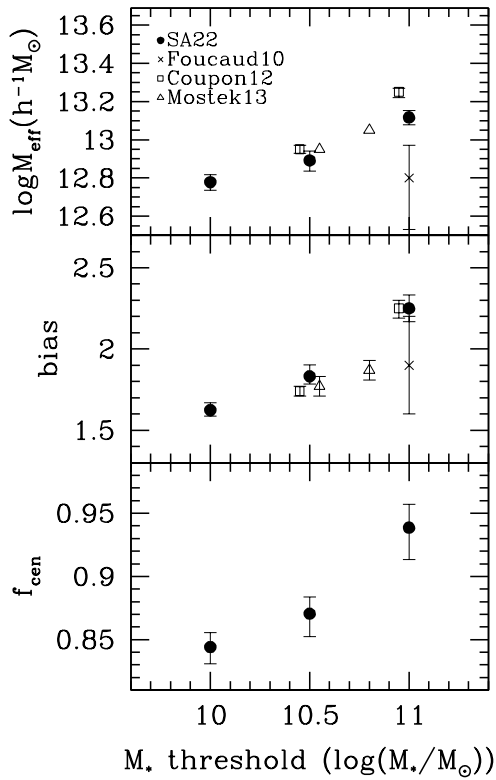}
\caption{Derived effective halo masses (top), biases (middle) and fractions of central galaxies (bottom) 
from the halo modeling. More massive galaxies reside in more massive haloes and tend to be 
central galaxies. As references, previous results from \citet{fou10}, \citet{cou12} and \citet{mos13} are also 
displayed. For the display purpose, literature values are slightly shifted on the $M_{*}$ axis.\label{figpar}}
\end{figure}

Figure~\ref{figpar} shows derived quantities as a function of stellar mass thresholds with filled circles. The 
top and middle panels show the effective halo mass and the bias 
for each sub-sample. Clearly massive galaxies tend to reside in massive haloes with high bias values, 
consistent with previous work. From the bottom panel of Figure~\ref{figpar}, we can see 
that massive galaxies also tend to be the central galaxy in a massive halo.

Recently, many researchers have measured the clustering of galaxies split by stellar mass.
For instance, \citet{fou10} measured 
the clustering of galaxies from the Palomar Observatory Wide-field Infrared Survey (Conselice 
et al. 2007). Their results show a halo mass of $\sim10^{12.8} h^{-1}M_{\odot}$ and a bias of 1.9 for 
$10^{11}M_{\odot}<M_{*}<10^{12}M_{\odot}$ galaxies at $0.8<z<1.2$, which is lower than ours (cross in Figure~\ref{figpar}). However, 
their analysis was done over a 1.5 deg$^{2}$ area, which is $>$5 times smaller than ours. In addition, they measured halo masses based on the 
``one galaxy per halo'' assumption with a correction for the halo occupation. 

As mentioned above, HOD analysis may be affected by cosmic variance if the survey area is too small
and the quality of the photometric redshifts used is poor.
With these issues in mind we quote two more results, one from a much wider area and another one based on the spectroscopic 
information. First, \citet{cou12} performed the halo modeling to reproduce the clustering of galaxies 
categorized by luminosity and type from the full CFHTLS--Wide survey area ($\sim133$ deg$^{2}$). 
Of our sub-samples, galaxies with $M_{*}>10^{10.5}M_{\odot}$ and $10^{11.0}M_{\odot}$ have similar number 
densities to their sub-samples containing all galaxies at $M_{g}-5\rm log h<-20.8$ and $<-21.8$ at $0.8<z<1.0$. 
Their estimates were $M_{eff}$ of $10^{12.95} h^{-1}M_{\odot}$ and $10^{13.25} h^{-1}M_{\odot}$, bias 
of $1.74$ and $2.25$, and the satellite fraction ($f_{sat}$) of $0.13$ and $0.06$ for 
$M_{g}-5\rm log h<-20.8$ and $< -21.8$, respectively. Overall, these results are in good agreement with our 
estimates, although the halo mass for the brightest luminosity bin is about 0.15 dex higher than ours 
(open squares in Figure~\ref{figpar}).
Second, \citet{mos13} used galaxies with spectroscopic information from the DEEP2 Galaxy 
Redshift Survey \citep{new13}
to measure a projected correlation function of galaxies at $0.74<z<1.4$ at different stellar masses, 
SFRs and sSFRs. They estimated bias and mean halo mass as ($1.77$, $10^{12.95} h^{-1}M_{\odot}$) 
and ($1.87$, $10^{13.05} h^{-1}M_{\odot}$) for all galaxies at $0.74<z<1.05$ with 
$M_{*}>10^{10.5}$ and $10^{10.8}M_{\odot}$, respectively (open triangles in Figure~\ref{figpar}). Although the 
stellar mass thresholds are slightly different from our samples, all estimates are in good agreement with our values.

It is interesting to note that our estimates are consistent with CFHTLS (wide-area survey) and DEEP2 (a wide spectroscopic survey), 
while small area surveys give somewhat different results. \citet{kim11} demonstrated the importance of 
surveyed area for the reliable measurement of galaxy clustering on large scales in overcoming cosmic variance 
studying Extremely Red Objects. The scatter of clustering on large scales is $\sim$30\% and $\sim$20\% in areas of 0.26 deg$^{2}$ and 
0.6 deg$^{2}$, respectively.
In addition, the halo model with HODs may be a more reliable scheme to measure the masses of host haloes for all, central and satellite galaxies, 
since this takes into account a more realistic contribution of central and satellite galaxies. Based on 
HODs discussed in this section, we can extend our analysis for galaxies split into stellar mass bins, 
which is an aspect that has not been fully explored in the literature.

\subsection{{\it Mass Bin}}\label{mbin}

\begin{figure*}
\centering
\includegraphics[width=100mm]{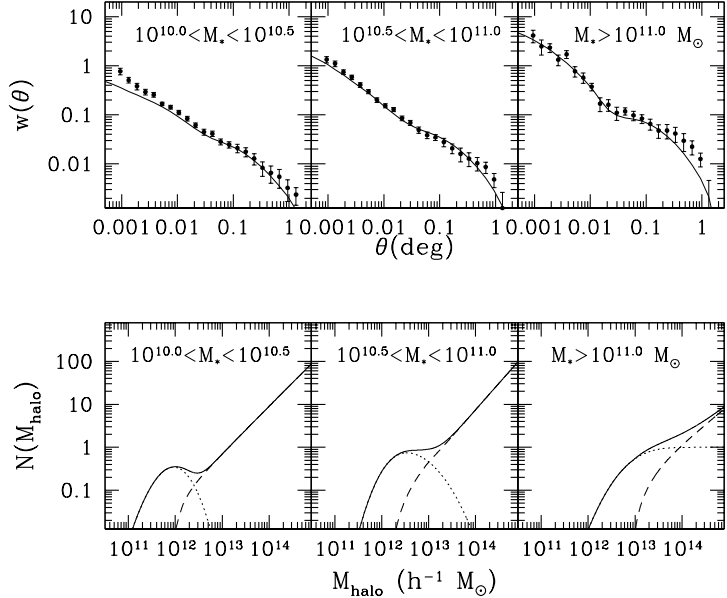}
\caption{The similar plot with Figure~\ref{figmth} for stellar mass binned galaxies. All symbols are 
the same to Figure~\ref{figmth}. However, the solid lines in the upper panel are reproduced angular 
correlation functions from the calculated HODs based on those for mass threshold galaxies (see text 
for more details). The most massive bin is identical to the right panel in Figure~\ref{figmth}. 
More massive central galaxies are more widely distributed in halo masses than less 
massive ones.\label{figmbin}}
\end{figure*}

If we are able to find the HODs of galaxies in various stellar mass bins, we can understand the relation 
between galaxies and their host haloes. However it is 
difficult to model the clustering of galaxies in stellar mass bins, since a different HOD 
shape should be assumed that is not well understood yet. In this section, we use the HODs discussed 
in \S~\ref{mth} to obtain the HODs of galaxies in different stellar mass bins.

\citet{zeh11} took the difference of HODs of luminosity threshold samples, in order to obtain the 
HODs of luminosity binned samples. We adopt this method to obtain the HODs of galaxies in different 
stellar mass bins. For instance, we obtain a HOD of galaxies at $10^{10.0}M_{\odot}<M_{*}<10^{10.5}M_{\odot}$ 
by subtracting the HOD of $M_{*}>10^{10.5}M_{\odot}$ galaxies from the HOD of $M_{*}>10^{10}M_{\odot}$ galaxies. The bottom 
panels of Figure~\ref{figmbin} show the HODs for $10^{10.0}M_{\odot}<M_{*}<10^{10.5}M_{\odot}$ 
(left) and $10^{10.5}M_{\odot}<M_{*}<10^{11.0}M_{\odot}$ (middle) galaxies, respectively, estimated in this way. For 
comparison, we also plot the HOD for $M_{*}>10^{11.0}M_{\odot}$ galaxies in the right panel. The upper panels 
display the observed angular correlation functions (points) for each stellar mass binned sample. The solid lines in the top-left 
and top-middle panels are not the best fit results, but the modeled clustering based on the stellar mass binned HODs. The top-right 
panel is identical to that in Figure~\ref{figmth}. 
The modeled correlation functions show a good agreement with the observed ones, except at a very small scales ($\theta\sim0^{\circ}.001$) 
where the modeled correlation functions are slightly underestimated in comparison to the observed ones.

Returning to the HODs, we note that the HOD for central galaxies is similar to a log--normal distribution, which indicates that 
there is a crude correlation between the host halo mass and the stellar mass of the central galaxies \citep{mos10}. The width of the 
distribution reflects the amount of the scatter in this relation \citep{zhe05}, and we find that the central HOD for more 
massive galaxies covers a wider halo mass range at a given value of $N_{c}(M_{halo})$ than that for 
less massive ones with the peak position shifted toward the 
high halo mass regime. As shown in the previous section, 
more massive galaxies show a gentler central HOD shape than less massive ones, which 
means a larger scatter between the stellar mass of central galaxies and the host halo mass. 
Therefore the broader shape of the HOD of massive galaxies reflects this effect. This can be caused by 
the stellar mass growth being stopped by any quenching mechanism, whilst the haloes keep growing.
For instance, \citet{gon11} pointed out that the inclusion 
of AGN feedback leads to a change in slope and a larger scatter of the relation between $K$-band luminosities 
of central galaxies and their host halo masses, since the feedback presents gas cooling. In addition, \citet{con13} 
compared the HODs predicted by different semi-analytic models and show that the central HODs of 
more massive galaxies are more affected by this feedback in all simulations. The idea is that above a certain luminosity 
or mass, galaxies do not grow to have a very large stellar mass due to the negative feedback by AGN, resulting 
in cases where very massive haloes possess central galaxies with reduced (but still massive) stellar masses. 
Consequently, the HOD of massive galaxies becomes extended toward a large halo mass.

\begin{figure}[h]
\includegraphics[scale=0.6]{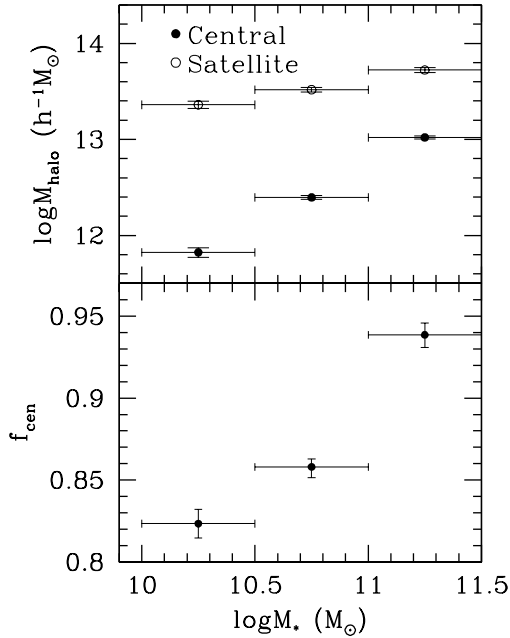}
\caption{The upper panel shows the derived halo masses for central (filled circles) and satellite galaxies (open circles) 
in different stellar mass bins. The bottom panel represents the fraction of central galaxies. For the 
highest stellar mass bin, we use the $M_{*}>10^{11}M_{\odot}$ galaxies. The horizontal bars indicate 
the size of stellar mass bins.\label{figparbin}}
\end{figure}

The upper panel of Figure~\ref{figparbin} shows the effective mass of dark matter haloes hosting galaxies 
in each stellar mass bin calculated from the HODs. For the most massive bin, we plot the 
stellar mass threshold sample of $M_{*}>10^{11}M_{\odot}$. The mass of dark matter haloes hosting 
central galaxies (filled circles) increases as the stellar mass increases with close to a linear slope. 
That for satellites (open 
circles) also shows a similar trend, but not as dramatic as it is for central galaxies. 
The reason for this difference in satellite galaxies is that a massive halo contains a 
large number of satellites 
in addition to a massive central galaxy. 
The bottom panel of Figure~\ref{figparbin} displays the fraction of central galaxies in each stellar 
mass bin. This indicates that the massive galaxies are more likely to be central galaxies
as the halo mass increases.

\subsection{{\it Stellar Mass to Halo Mass Ratio}}\label{mratio}

Since galaxies evolve in dark matter haloes, the properties of galaxies depend on their host 
dark matter haloes. In this context, the relation between stellar mass ($M_{*}$) and halo mass ($M_{\rm halo}$) 
is a good testbed to constrain the evolution of galaxies. In addition, 
the ratio between the stellar mass of central galaxies and the halo mass is sensitive to the 
conversion efficiency from baryons to stellar mass in the central galaxy \citep{zhe07}. 
In previous work, the $M_{*}$--$M_{\rm halo}$ (or $M_{*}/M_{\rm halo}$--$M_{\rm halo}$) relation was derived  using
several different methods such as HOD, Conditional Luminosity Function and Sub-halo Abundance Matching 
\citep{zhe07,beh10,mos10,wan10,wak11}. However, one of the advantages of the HOD framework is that the clustering and 
the number density of galaxies are fitted simultaneously without any assumption about the scatter between $M_{*}$ and 
$M_{\rm halo}$ being a fixed value. Here, we derive the relation based on the best fit 
HODs directly, unlike previous HOD work, fitting a functional form to best fit HOD parameters.

We use the HODs for stellar mass binned samples discussed in the previous section to calculate the 
stellar mass to halo mass ratio ($M_{*}/M_{\rm halo}$). Since the HOD is the mean number of galaxies at a given halo mass, 
the summation of HODs multiplied by mean stellar masses for a given stellar mass bin represents the 
stellar mass at a given halo mass. Therefore, the stellar mass at a 
given halo mass can be calculated by 
\begin{equation}
M_{*}(M_{\rm halo}) = \sum N_{i}(M_{\rm halo})\langle M_{*} \rangle_{i}
\end{equation}
where $N_{i}(M_{\rm halo})$ is the central or satellite HOD for the $i$th stellar mass binned sample and 
$\langle M_{*} \rangle_{i}$ is 
the mean stellar mass of galaxies in the $i$th stellar mass bin. In this work, there are three 
stellar mass bins, each corresponding to $10^{10}M_{\odot}<M_{*}<10^{10.5}M_{\odot}$, 
$10^{10.5}M_{\odot}<M_{*}<10^{11.0}M_{\odot}$ and $M_{*}>10^{11}M_{\odot}$. 
Using this stellar mass, we calculate the stellar mass to halo mass ratio.

\begin{figure}[h]
\centering
\includegraphics[width=54mm]{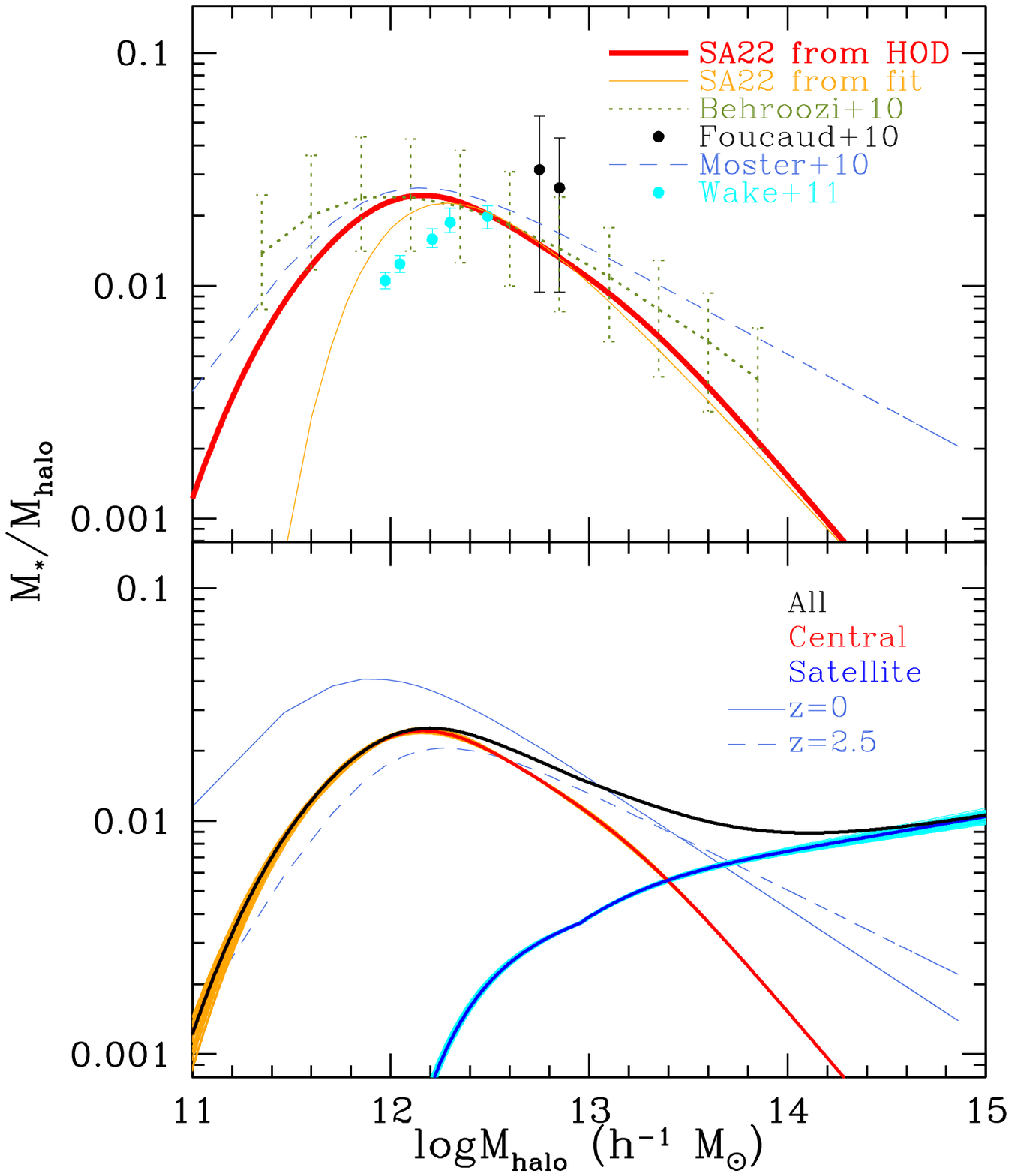}
\caption{The stellar mass to halo mass ratio based on the best fit HODs. The upper panel is for 
central galaxies only at $z\sim1$. The result of this work is represented by a thick red line. For the comparison, we also plot 
previous results at $z\sim1$ from \citet{beh10} (green dotted line), \citet{fou10} (black points), \citet{mos10} (blue short dashed line) 
and \citet{wak11} (cyan points). 
The ratio shows a peak at $\rm log M_{\rm halo}\sim 10^{12} h^{-1}M_{\odot}$, which is consistent with 
previous results. 
The orange line is the ratio calculated by the empirical function in \citet{wak11} for our result. 
The lower panel displays the ratio 
for central (red) and satellite (blue) galaxies. The black line is the combination of both central and 
satellite galaxies. The orange and cyan lines are results with varying HOD parameters randomly within the uncertainty range 
for central and satellite galaxies, respectively. Solid and dashed lines are from \citet{mos10} for $z=$0 and 2.5 respectively. 
Satellites mainly contribute to the total stellar mass in group or cluster scale haloes. 
\label{figmsmh}}
\end{figure}

The upper panel in Figure~\ref{figmsmh} shows the stellar mass to halo mass ratio for the central galaxy. 
The red solid line is the result based on equation (8). 
The peak of the ratio is located at $\sim10^{12.16} h^{-1}M_{\odot}$ with a ratio of 0.024. Below and above 
this halo mass, the ratio drops rapidly. This means that the conversion from baryons to stellar 
mass in the central galaxy is the most efficient in $\sim10^{12} h^{-1}M_{\odot}$ dark matter haloes as
traced at $z\sim1$. For the comparison, we also display the ratio at $z=1$ (royal blue line) from \citet{mos10}. 
We have to note that the ratio may be underestimated for low mass haloes in this work, since the stellar mass 
limit is $M_{*}=10^{10}M_{\odot}$, and less massive galaxies are missed. Therefore the actual 
slope of the ratio in the low halo mass regime may be flatter than our measurement.

Here, we also compare our result with other previous works. \citet{wak11} also performed the halo modeling for stellar 
mass limited samples from the NEWFIRM Medium Band Survey \citep{bra09,vand09,vand10,whi11}. 
After fitting the empirical function in \citet{zeh11} to the best fit HOD parameters, 
they obtained the ratio at $z=1.1$ (cyan points in Figure~\ref{figmsmh}). 
Their result shows a peak at a higher halo mass, and a flatter (steeper) shape in a high (low) 
halo mass regime than ours, although the relation is poorly constrained at high and low halo masses. If we also fit the same function to our best fit parameters, 
this also shows a steeper slope in the low halo mass regime, but the discrepancy is not so significant 
in the high mass regime (orange line). On the one hand, this means that the functional form may underestimate 
the ratio for less massive haloes, since this does not fully take into account HODs for stellar mass binned 
samples. On the other hand, for the high halo mass regime, the flatter trend may be caused by the different 
parameter set for HODs as \citet{wak11} fixed the transition shape of central HODs with $\sigma_{cut}=0.15$. In fact, 
the relation between derived halo mass parameters for central galaxies ($M_{cut}$ in equation (3)) and stellar 
masses in 
\citet{wak11} is flatter than ours. If we make the relation for our fitted parameters flatter arbitrarily, the ratio becomes much 
flatter in the high halo mass regime. \citet{fou10} also estimated the ratio as 0.032$\pm$0.022 and 
0.026$\pm$0.017 for 
$10^{11}M_{\odot}<M_{*}<10^{11.5}M_{\odot}$ and $10^{11.5}M_{\odot}<M_{*}<10^{12.0}M_{\odot}$ at 
$0.8<z<1.2$, 
respectively (black points). Although these are in agreement within the uncertainty range, their halo masses 
are smaller compared to ours as mentioned above. Therefore it is possible 
that their points move toward a lower ratio and a higher halo mass regime, and become 
consistent with ours.

Additionally, \citet{beh10} obtained the same quantity using the abundance matching technique. We plot their 
result for the central galaxy including full uncertainties at $z=1$ in the upper panel of 
Figure~\ref{figmsmh} (green points and line). Our result agrees with theirs within the uncertainty. 
However, our result may be slightly steeper in both the high and low halo mass regimes than the \citet{beh10} result. 
We find 
a similar discrepancy from \citet{mos10} as well (royal blue). As mentioned above, the ratio in the low 
halo mass regime is underestimated due to the stellar mass limit in this work. However,
in the high halo mass regime, that is not the case and other factors may account for this
discrepancy. First, \citet{beh10} and \citet{mos10} included the scatter in the $M_{*}$-$M_{\rm halo}$ relation 
with a fixed value. Whereas, our HOD takes into account such a scatter as shown in Figure~\ref{figmth}.
\citet{beh10} pointed out that a larger scatter in the relation  
makes the relation steeper. Second, the stellar mass uncertainty on individual galaxies can also affect the 
result, since the number of low mass galaxies having overestimated stellar masses is larger than that of 
high mass galaxies with underestimated stellar masses. If this scatter is included, the relation also becomes steeper 
\citep{beh10}. 

In addition to the central galaxy, satellites also contribute to the total baryons in dark matter 
haloes. \citet{beh10} and \citet{fou10} pointed out that satellites account for the majority of the 
total stellar mass in more massive dark matter haloes. As we discussed in the previous section, 
satellite galaxies with $M_{*}>10^{10}M_{\odot}$ tend to live in massive haloes.
The bottom panel in Figure~\ref{figmsmh} also shows this trend. The red line is the 
ratio for the central galaxy only, which is the same as in the upper panel. The blue line is for satellites 
calculated by equation (8) with the HODs in Figure~\ref{figmbin}. However, in this case, 
the value represents the ratio of the total stellar mass in satellites to the total mass of the dark matter halo. 
Orange and cyan lines are the results for central 
and satellite galaxies, respectively, which are derived 
from 100 different HOD sets with varying parameters ($\sigma_{cut}, M_{1}$ and $\alpha$) with  values varied randomly within
their measured uncertainty ranges.
In contrast to
the central galaxy, the ratio for satellites increases with increasing halo mass. In addition, the $M_{*}/M_{\rm halo}$ 
ratio for satellites is comparable with the value for the central galaxy in group environments 
($10^{13} h^{-1}M_{\odot}<M_{\rm halo}<10^{14} h^{-1}M_{\odot}$) 
and dominant in cluster environments ($M_{\rm halo}>10^{14} h^{-1}M_{\odot}$). As discussed in \citet{beh10} and 
\citet{fou10}, the accretion of dark matter toward a massive halo leads to the rapid growth of the halo 
mass. At the same time, infalling galaxies become satellites and contribute to the total stellar mass. 
However, the growth of central galaxies is not so efficient in this regime 
due to the deep potential well preventing an efficient merger or cold gas accretion onto the central galaxy. We also note that 
our data do not include faint, possibly numerous satellite galaxies.
Thus, the $M_{*}/M_{\rm halo}$ ratio for satellites may be underestimated and can only be addressed with deeper data.

From the $M_{*}/M_{\rm halo}$ ratio for the central galaxy and satellites, we find that 
the star formation efficiency for the central galaxy is the most efficient in 
$\sim10^{12} h^{-1}M_{\odot}$ dark matter haloes at $z=1$ with the $M_{*}/M_{\rm halo}$ ratio peaking at about 0.02. Moreover, 
we show that satellites are the dominant contributor to the $M_{*}/M_{\rm halo}$ of high mass haloes ($M_{\rm halo}>10^{13.5}h^{-1}M_{\odot}$). 
When we consider both central and satellite galaxy samples with $M_{*}>10^{10}M_{\odot}$, we find the star formation efficiency in terms of 
total halo mass and stars in all galaxies is 1\%--2\% consistently. However, the cut--off at low halo mass is created by our 
selection of massive galaxies, so the mass of stars is not fully accounted for in the lowest mass haloes.
We also confirm the evolution of this relation, comparing to the result at different redshift from \citet{mos10} 
(the bottom panel in Figure~\ref{figmsmh}).



\section{CLUSTERING WITH STAR-FORMING ACTIVITY}\label{clssfr}

The clustering of galaxies has been shown to depend on star formation rate and color. 
In this section, we split galaxies at $0.8<z<1.2$ based on their star forming activity to investigate how their clustering properties 
depend on star formation. Additionally, since sSFR is roughly related to the star formation 
efficiency in a galaxy, we use various sSFR criteria to define each sub-sample.

\subsection{{\it Passive vs. Star Forming}}\label{pgsf}

We start with the comparison of bias values for different populations of galaxies. For this analysis, galaxies are 
split into passive and star-forming. In order to define the sSFR cut for passive galaxies, we check galaxies detected by 
{\it Spitzer} from the IRSA catalog\footnote{http://irsa.ipac.caltech.edu/applications/Gator/}. We cross-match 
{\it Spitzer} sources to our galaxies with $M_{*}>10^{10}M_{\odot}$ and $0.8<z<1.2$, and find 1871 IRAC and 213 24$\mu$m 
sources. Of 24$\mu$m sources, 15\% and 8\% satisfy $\rm log\,sSFR/yr^{-1}<-10$ and $-10.5$, respectively. This means that 
the sSFR cut may not be a clean way 
to classify pure passive galaxies, but we can isolate them efficiently. Thus, 
although there are two galaxy loci in Figure~\ref{figmssfr} separated by 
$\rm log\,sSFR/yr^{-1}\sim-10$, we classify passive galaxies with $\rm log\,sSFR/yr^{-1}<-10.5$ in order to minimize the 
contamination by dusty galaxies. Then, star-forming galaxies are defined by $\rm log\,sSFR/yr^{-1}>-10$. 
we then apply stellar mass cuts with $M_{*}>10^{10.0}$, $10^{10.5}$ and $10^{11.0}M_{\odot}$ for each population. 
However, since the number of star-forming galaxies in the highest mass bin is too small to measure the clustering, we do not 
include this sub-sample in our analysis.

\begin{figure}[h]
\centering
\includegraphics[width=70mm]{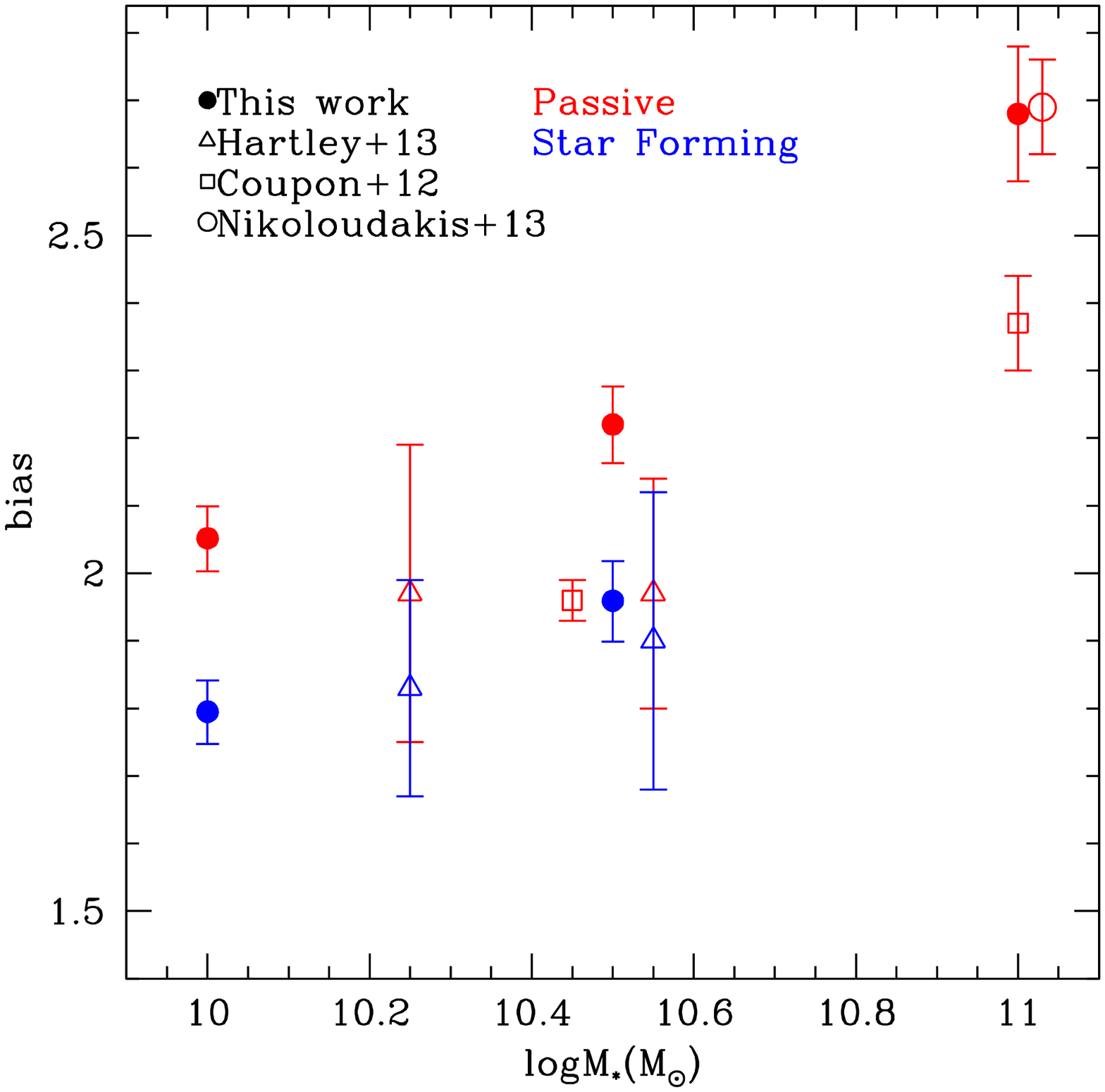}
\caption{Biases for passive (red) and star-forming (blue) galaxies in $\rm log M_{*}/M_{\odot}>10.0$, $10.5$ and $11.0$ 
(filled circles). Open triangles are from \citet{har13} with $10.0<\rm log M_{*}/M_{\odot}<10.5$ and $\rm log M_{*}/M_{\odot}>10.5$. 
Open squares are for red galaxies in \citet{cou12} with $M_{g}-5\rm log h<-20.8$ and $-21.8$. Also the open circle shows the result of 
LRGs at $z\sim1$ in \citet{niko13}. For the display purpose, open symbols are slightly shifted along the $M_{*}$ axis.\label{figpgsf}}
\end{figure}

For each selected sub-sample, the angular two-point correlation function is measured. Since the intrinsic 
HOD shape for each sample is not well understood, we estimate the bias by fitting the angular 
correlation function of the underlying dark matter instead of finding the best fit HODs. First, we obtain the 
real space correlation function of dark matter from the linear matter power spectrum at $z=1$. 
Then the real space correlation function is transformed into the angular correlation 
function with the observed redshift distribution, after multiplying it by the bias value. Through fitting 
the transformed angular correlation function to the observed one, we find the best fit bias values 
for each sub-sample. The fitting is performed to the angular range between 0.02$^{\circ}$ and 0.7$^{\circ}$, 
where the influence by the integral constraint is minimized and 2-halo component dominates.

Filled circles in Figure~\ref{figpgsf} show the estimated bias for passive (red) and star-forming (blue) galaxies.
It is clear that passive galaxies are more clustered than star-forming ones, which is consistent with previous results 
at the similar redshift \citep{coi08,mcc08,wil09,har10,har13,bie14}. Additionally, high mass galaxies show a
stronger clustering strength than low mass ones independently of population. On the other hand, some previous results pointed out 
that passive or red galaxies show similar (or higher) clustering strengths with decreasing stellar masses (DEEP2 in Coil et al. 2008 
and UKIDSS UDS in Williams et al. 2009 and Hartley et al. 2013).
Open triangles in Figure~\ref{figpgsf} are for galaxies with $10^{10}M_{\odot}<M_{*}<10^{10.5}M_{\odot}$ and $M_{*}>10^{10.5}M_{\odot}$ at 
$z\sim1.06$ from \citet{har13} based on the UKIDSS UDS data. Passive galaxies show a similar clustering strength in both bins, which 
implies that there is no stellar mass dependence. This was interpreted as the contribution of low mass satellite galaxies in massive haloes, 
leading to similar clustering strengths independently of the stellar mass. 

\begin{figure*}
\centering
\includegraphics[width=83mm]{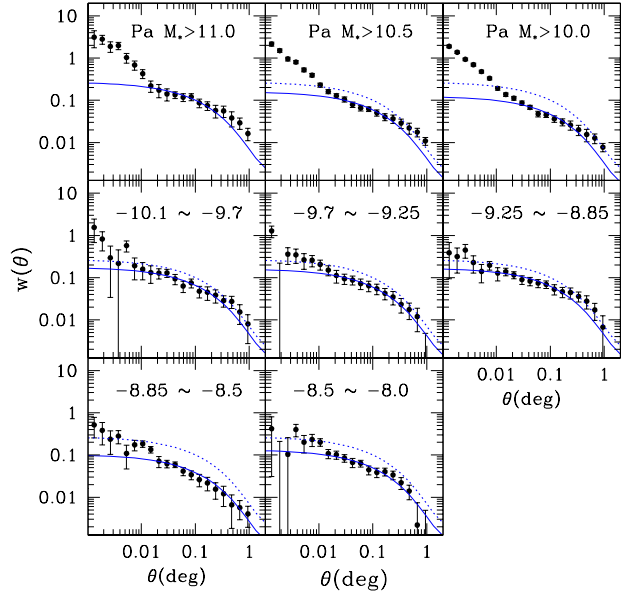}
\caption{Angular correlation functions (points) for sSFR binned samples. The top row is for passive 
galaxies ($\rm log\,sSFR/yr^{-1}<-10.5$) with different stellar mass thresholds. The middle and bottom rows are for 
star-forming galaxies in various sSFR bins as labeled. The solid line is the best fit result, and the dotted line 
is for $M_{*}>10^{11}M_{\odot}$ passive galaxies. Comparing solid and dotted lines, the amplitude difference is 
the largest in the bottom-left panel, which indicates the lowest bias.\label{figssfrfit}}
\end{figure*}

Our results may appear to be different from previous literature results. However, we also display bias values for red galaxies with 
$M_{g}-5\rm log h<-20.8$ and $-21.8$ (red open square) at $0.8<z<1.0$ from CFHTLS in 
\citet{cou12}. The number densities in each magnitude bin are similar to those for our passive galaxies with $M_{*}>10^{10.5}$ and $10^{11}M_{\odot}$, 
respectively. Although their measurements are lower than ours, brighter red galaxies are more clustered than fainter ones, which is the same trend 
to our passive galaxies. Possibly lower values are caused by different selection criteria, since absolute magnitude limited samples can include 
low stellar mass galaxies, but exclude some high mass ones. In addition, they noted that red galaxies at $z>0.8$ were contaminated by blue galaxies. 
The similar trend was also found in \citet{mcc08}, which showed early-type galaxies with brighter 
absolute magnitude are more strongly clustered than fainter early-type galaxies at $0.7<z<1.1$, based on the CFHTLS data. Finally, we also note the 
bias value from LRGs at $z\sim1$ from \citet{niko13} (red open circle). The number density of LRGs in their paper is similar to that of our 
high mass passive galaxies, and the estimated bias is also consistent.

Since the area coverage of our data is not wide enough to fully overcome the effect of cosmic variance, it may be hard to conclude whether the 
clustering strength of passive galaxies at $z\sim$1 is correlated with stellar mass in our dataset 
conclusively. 
However, our result shows the consistent trend with the result from the wider survey data, which indicates our dataset is not significantly affected 
by cosmic variance. Based on this conclusion, we investigate the clustering property of galaxies with finer sSFR bins below.

\subsection{{\it sSFR bins}}\label{ssfrbin}

As shown in Section~\ref{sedfit}, the main sequence of star-forming galaxies has a slope of 
$-0.33$ on the sSFR versus stellar mass plane. However, our main goal is to investigate the clustering strength as a function 
of star formation activities at $z\sim1$. Thus, we simply apply 
five sSFR bins for star-forming galaxies 
($-8.5<\rm log\,sSFR/yr^{-1}<-8.0$, $-8.85<\rm log\,sSFR/yr^{-1}<-8.5$, $-9.25<\rm log\,sSFR/yr^{-1}<-8.85$, 
$-9.7<\rm log\,sSFR/yr^{-1}<-9.25$ and $-10.1<\rm log\,sSFR/yr^{-1}<-9.7$). The width of each bin is determined to 
include sufficient galaxies for a reasonable clustering measurement. Additionally, 
passive galaxies are defined as galaxies with $\rm log\,sSFR/yr^{-1}<-10.5$, and they are further divided into mass 
thresholds of $M_{*}>10^{10}$, $10^{10.5}$ and 
$10^{11}M_{\odot}$. Column 3 in Table~\ref{tssfr} lists the number of galaxies in each bin. We also compare the 
clustering properties in each sSFR bin with narrower stellar mass bins later.

Figure~\ref{figssfrfit} displays the angular correlation function 
of each sub-sample (points) and the best fit result (solid line). The top three panels are for passive 
galaxies with different stellar mass thresholds. The second and third rows are results for star-forming galaxies 
in various sSFR bins with $M_{*}>10^{10}M_{\odot}$. The labels in the second and third rows indicate 
the sSFR range in a logarithmic scale. We also plot the best fit result for $M_{*}>10^{11}M_{\odot}$ 
passive galaxies with 
dotted line. After comparing the solid and the dotted lines for sSFR binned samples, the difference is the largest for 
$-8.85<\rm log\,sSFR/yr^{-1}<-8.5$ galaxies, which means it has the lowest bias among our sub-samples.
On small scales, we are able to find excess of clustering amplitudes 
from all sub-samples, although the measurement with relatively large uncertainties is affected by the small number of objects. 
However, the excess for star-forming galaxies looks less significant than passive ones. 
This may mean that the number of star-forming satellites in each bin is lower than passive satellites, which intrinsically weakens the clustering 
strength. However, the actual spatial distribution of satellites in each sub-sample may also influence the clustering. For an extreme example, 
if a star-forming sub-sample includes one satellite per halo and just a portion of centrals, the small scale clustering is more weakened than the large scale.
Unfortunately, our data is not enough in depth and area to demonstrate these effects separately (or simultaneously). Larger and deeper datasets 
in the future will allow us to investigate this issue. In this work, we will focus on the bias or halo mass estimated from the large scale clustering only.

\begin{figure}[h]
\centering
\includegraphics[scale=0.45]{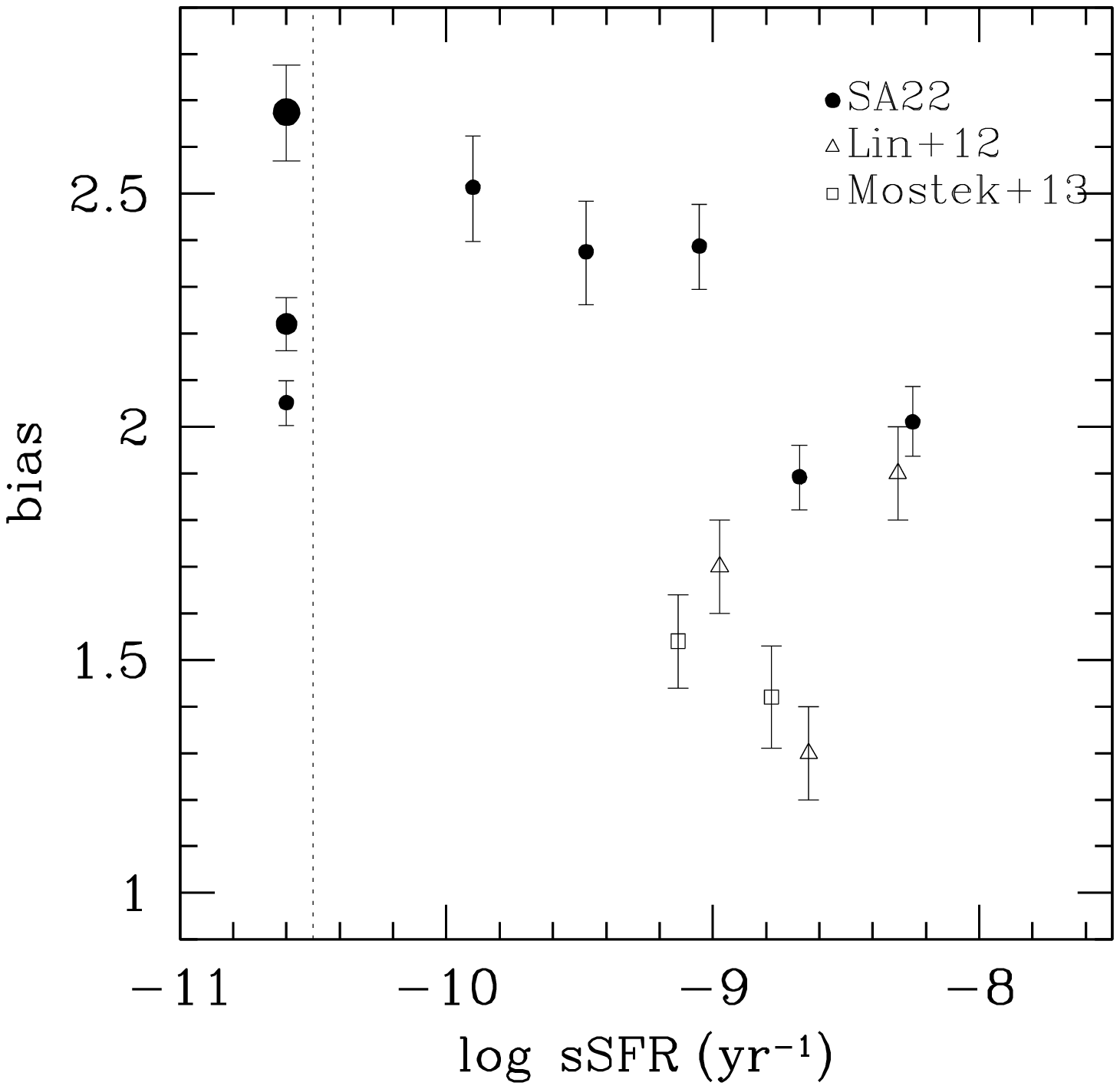}
\caption{The bias of sSFR binned galaxies. The vertical dotted line shows the criterion distinguishing star-forming 
and passive galaxies. For passive galaxies ($\rm log\,sSFR/yr^{-1}<-10.5$), a symbol size corresponds to stellar 
mass thresholds, $M_{*}>10^{10}$, $10^{10.5}$ and $10^{11}M_{\odot}$ from small to large points. The previous 
results for $z\sim2$ sBzKs \citep{lin12} and $z\sim1$ galaxies \citep{mos13} are also plotted with open triangles and open squares, respectively.
The clustering strengths increase with decreasing sSFRs. However, it is also found that the highest sSFR galaxies are more 
strongly clustered than main sequence galaxies.
\label{figbssfr}}
\end{figure}

Figure~\ref{figbssfr} shows the estimated bias as a function of sSFR (points). The size of the points 
for passive galaxies represents the stellar mass thresholds which are $M_{*}>10^{10}$, $10^{10.5}$ and 
$10^{11}M_{\odot}$ from small to large points, respectively. For star-forming galaxies, a single mass threshold of 
$M_{*}>10^{10}M_{\odot}$ is applied. For comparison, we also plot the result of blue galaxies with 
$M_{B}<-20.5$ at $0.74<z<1.05$ from \citet{mos13} (open squares).
As found by \citet{mos13}, we also confirm that the bias or a clustering strength 
decreases with increasing sSFRs up to $\rm log\,sSFR/yr^{-1}\sim-8.6$. The discrepancy of biases may come from 
different selection criteria, since they have used rest-frame magnitude limited samples that may include $M<10^{10}M_{\odot}$ galaxies
which can dilute their measured clustering strength. Since the sSFR value for the 
main sequence at $z=1$ is $\rm log\,sSFR/yr^{-1}\sim-9$ in Figure~\ref{figmssfr}, galaxies just above (or the upper part of) the main sequence 
show the lowest bias, which is consistent with \citet{mos13}. This anti-correlation has been 
reported at $z\sim1$ by \citet{coi08} and \citet{mos13}, and in the local Universe by \citet{li08} and \citet{hei09}. 
However, we also see a possible reversal of the relation in the highest sSFR bin ($-8.5<\rm log\,sSFR/yr^{-1}<-8.0$).
This was already noted by \citet{lin12} for sBzK galaxies at $z\sim2$, but we find a similar trend at $z\sim1$ and we plot their results 
with open triangles. Direct comparison of the absolute values is difficult due to different stellar 
mass limits. Galaxies in this work are more massive than those in \citet{lin12}. However, both results 
show the same trend, and we find a similar sSFR for the lowest bias sample. 

\begin{figure*}
\centering
\includegraphics[width=110mm]{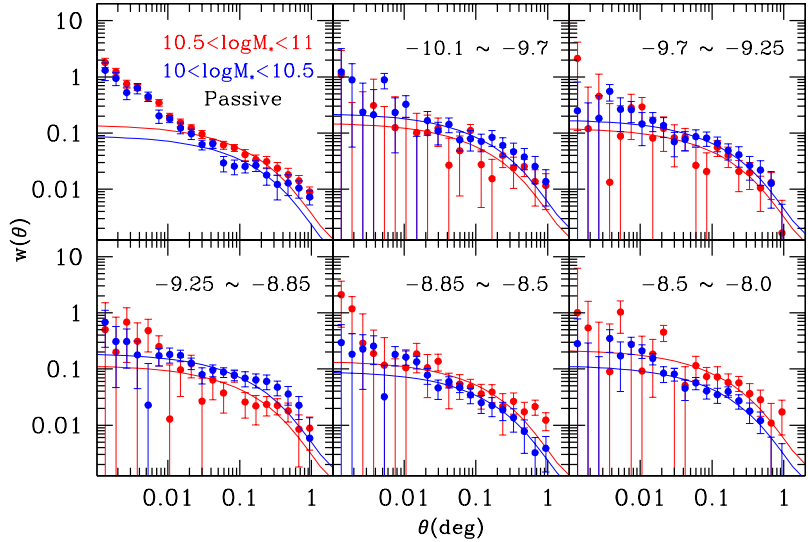}
\caption{Angular correlation functions for sSFR binned samples in $10^{10}<M_{*}/M_{\odot}<10^{10.5}$ (blue) and 
$10^{10.5}<M_{*}/M_{\odot}<10^{11}$ (red). Points with error bars are correlation functions measured and solid lines are 
the best fit dark matter clustering on large scales. Black labels in each panel indicate the sSFR range in a logarithmic scale.\label{figssfrbin}}
\end{figure*}

Additionally, we note the mass of haloes hosting each sub-sample. The halo mass is calculated using the measured 
bias and the halo mass function in \citet{tin10}. Since this procedure does not consider the inclusion of 
satellites, the calculated halo mass may show a discrepancy from that from the full HOD framework in previous sections. 
However, since our main goal here is to examine how the bias factor or host halo mass change as a function of sSFR, 
we will defer an HOD analysis to a future paper where we will include more galaxies over a wider area.
Details of the influence of satellites are well described in \citet{zhe07}. In Appendix~\ref{app2}, we also compare 
halo masses derived by the halo model and the direct fit. Our measured biases and halo masses are given in Table~\ref{tssfr}. 
We find that the mass of host haloes of galaxies with the lowest bias is 
$M_{\rm halo}=10^{12.684} h^{-1}M_{\odot}$. Also the mean halo masses for passive galaxies are 
$M_{\rm halo}=10^{12.817}$, $10^{12.940}$ and $10^{13.207} h^{-1}M_{\odot}$ 
from the lowest to the highest stellar mass bins, respectively. 

From Figure~\ref{figbssfr}, we show the clustering strength as a function of sSFRs. Now we investigate 
the same quantities using galaxies in different stellar mass bins, in order to check the influence of 
stellar mass as the negative main sequence slope on the sSFR--mass plane could arise from a selection bias toward high sSFR galaxies that are in
less massive haloes and hence are less clustered. 
To assess the level of this potential bias we further split galaxies into different stellar mass bins of 
$10^{10}M_{\odot}<M_{*}<10^{10.5}M_{\odot}$ (low mass, LM) and $10^{10.5}M_{\odot}<M_{*}<10^{11}M_{\odot}$ 
(high mass, HM) with the same sSFR bins mentioned above. 
Unfortunately, it is hard to measure the secure angular 
correlation function of star-forming galaxies with $M_{*}>10^{11}M_{\odot}$ divided into various sSFR bins because of the small 
number of objects. Thus, we do not consider the highest stellar mass bin here.
Figure~\ref{figssfrbin} shows the observed correlation function of sSFR binned samples in LM (blue) and HM (red). Symbols are the same 
as Figure~\ref{figssfrfit}. Black labels indicate sSFR ranges. As already shown in \S~\ref{pgsf}, passive galaxies in HM is more 
clustered than those in LM. On the other hand, star-forming galaxies with $\rm log\,sSFR\lesssim-9$ in LM show stronger clustering 
strengths than HM ones, but this is opposite at the high sSFR range.

\begin{figure}[h]
\centering
\includegraphics[scale=0.35]{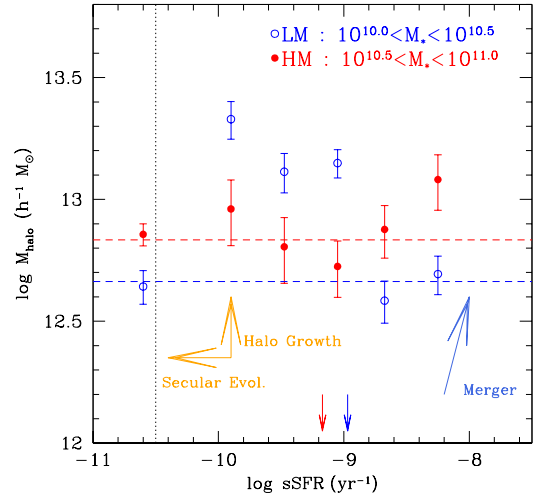}
\caption{The similar plot with Figure~\ref{figbssfr} for $10^{10}M_{\odot}<M_{*}<10^{10.5}M_{\odot}$ (blue open circle) 
and $10^{10.5}M_{\odot}<M_{*}<10^{11}M_{\odot}$ (red filled circle) galaxies. In this plot, we quote 
calculated halo masses instead of bias values. We confirm the trend in Figure~\ref{figbssfr} from narrower 
stellar mass binned samples. The arrows present possible evolutionary paths. 
See text for more details and discussion. The horizontal dashed lines indicate $M_{eff}$ values of haloes 
for all galaxies in each stellar mass bin. Downward arrows at the bottom are central sSFR values of the main sequence for 
each stellar mass bin.\label{figmssfrmb}}
\end{figure}

Figure~\ref{figmssfrmb} displays the halo mass as a function of sSFR for the 
$10^{10}M_{\odot}<M_{*}<10^{10.5}M_{\odot}$ (blue) and $10^{10.5}M_{\odot}<M_{*}<10^{11}M_{\odot}$ (red) 
bins, respectively. In this case, we plot the halo mass on the y--axis rather than the bias factor.
From both stellar mass bins, we confirm the anti-correlation 
in the sSFR--$M_{\rm halo}$ relation in the low sSFR regime of star-forming galaxies and 
at lower significance the reversal of the relation 
at $\rm log\,sSFR/yr^{-1}>-8.5$ shown in Figure~\ref{figbssfr}.
This means that the feature shown in Figure~\ref{figbssfr} is mainly caused 
by an environmental effect connecting the halo occupation of high sSFR galaxies to the mass of their host halo
and not due to an effect intrinsic to their stellar mass. One difference between two mass bins is the sSFR value 
showing the lowest halo mass. Although the uncertainty is substantial, HM star-forming galaxies with the lowest halo mass 
have lower sSFR than LM ones. Considering the slope for main sequence galaxies in Figure~\ref{figmssfr}, galaxies on the upper envelope of the main sequence 
in each stellar mass bin reside in the lowest mass haloes. 
The number of galaxies in each bin and all the measured values are listed in 
Table~\ref{tssfrbi}.

Finally, it is worth comparing our results to the known sSFR--local density relation. First, \citet{coo08} found the same 
anti-correlation at $z\sim1$ over the similar sSFR range from the DEEP2 survey. In addition, we can also see a suggestion of a weak reverse relation 
in the highest sSFR bin (see Figure 9 in their paper). Second, \citet{ko12} also reported the similar trend at $z=0.087$ 
based on the AKARI North Ecliptic Pole (NEP) Wide survey and A2255 from the AKARI CLusters of galaxies EVoLution studies (CLEVL; Im et al. 2008; 
Lee et al. 2009a). 
Although \citet{ko12} focused on the very low redshift, there is no significant difference in the bias--sSFR relations at low 
and high redshifts as we already mentioned above. Moreover, their field coverage is wide enough to study the relation from low to high 
local densities. In conclusion, our results also show good agreement with local density studies.

\section{DISCUSSION}\label{discuss}

In the previous section, we found a clear dependence of clustering strength on sSFR. Here, we 
further investigate the property of galaxies in the highest sSFR bin, which are in denser environments than 
main sequence star-forming galaxies,  and we present a possible scenario for the evolution of star-forming galaxies.

\begin{figure}[h]
\centering
\includegraphics[scale=0.55]{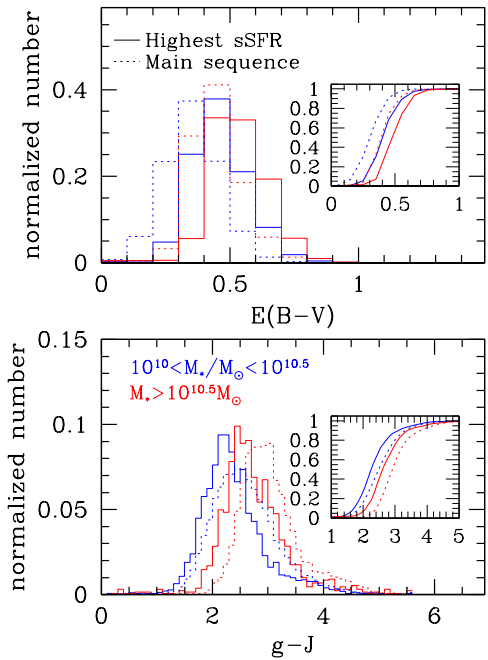}
\caption{$E(B-V)$ (top) and $g-J$ color (bottom) distributions of galaxies with $10^{10}<M_{*}/M_{\odot}<10^{10.5}$ (blue) and 
$M_{*}>10^{10.5}M_{\odot}$ (red).
Solid and dotted histograms are for the highest sSFR galaxies ($-8.5<\rm log\,sSFR/yr^{-1}<-8$) 
and main sequence ones ($-9.5<\rm log\,sSFR/yr^{-1}<-8.5$), respectively. The inset shows the cumulative distribution with the same 
line style to the main panel. The highest sSFR galaxies are more heavily 
obscured and show relatively bluer colors than main sequence galaxies at a given stellar mass.\label{figebvcol}}
\end{figure}

We begin with the comparison of the internal dust attenuation of $E(B-V)$ derived from SED fitting. The top panel of Figure~\ref{figebvcol} 
shows the $E(B-V)$ distribution of the highest sSFR galaxies ($-8.5<\rm log\,sSFR/yr^{-1}<-8.0$) and the main sequence galaxies 
($-9.5<\rm log\,sSFR/yr^{-1}<-8.5$) with solid and dotted histograms, respectively. In addition, galaxies are also split into two stellar mass bins 
of $10^{10}<M_{*}/M_{\odot}<10^{10.5}$ (blue) and $M_{*}>10^{10.5}M_{\odot}$ (red). The inset shows the cumulative distribution of each 
sub-sample. The median values of $E(B-V)$ for the highest sSFR bin is 0.45 and 0.55 for 
$10^{10}<M_{*}/M_{\odot}<10^{10.5}$ and $M_{*}>10^{10.5}M_{\odot}$ galaxies, respectively. However, those for main sequence galaxies in the same stellar mass bins 
are 0.375 and 0.45, respectively.
It is clear that high sSFR galaxies are more heavily obscured at a given stellar mass and in strongly biased regions than the main sequence galaxies.
\citet{whi12} categorized high sSFR galaxies as dusty, blue star-forming galaxies, and noted that they are possibly merger driven starburst galaxies.
We also show in Figure~\ref{figebvcol} the distribution of $g-J$ color (bottom) corresponding to the rest-frame UV--optical color. 
The line styles are the same as the top panel. This bluer color is also confirmed by the best fit SED in Figure~\ref{figsedexam}. 
The bottom panels in Figure~\ref{figsedexam} are examples of high sSFR galaxies. Compared to main sequence galaxies (middle panels in Figure~\ref{figsedexam}), it is clear that 
high sSFR galaxies show significantly higher fluxes in the rest-frame UV regime. 
Although there is a significant overlap, high sSFR galaxies 
are either relatively blue $g-J$ color or have a higher extinction suggesting the presence of an obscured young stellar population.

Although these high sSFR galaxies are most likely  
to be obscured galaxies with efficient star formation, 
their colors could also be strongly affected by the presence of an AGN.
From the 
previous work based on the clustering, AGN reside in dark matter haloes of $>10^{13}h^{-1}M_{\odot}$ (e.g., Hickox et al. 2009, 2011; Ross et al. 2009b; 
also see a review in Cappelluti et al. 2012 and references therein). Interestingly, this halo mass is consistent to our result for high sSFR galaxies. 
Therefore, in order to test the possibility of AGN contamination, we check the fraction of AGN in this high sSFR bin. 
Using {\it Spitzer} sources mentioned above, we select 75 potential AGN with IRAC colors suggested by \citet{ste05}. Among these AGN, 15 sources 
are in the highest sSFR bin, and this corresponds to $\sim$8.8\% of all IRAC sources matched with our galaxies in the same sSFR bin. 
In addition to {\it Spitzer} data, we also select potential AGN in the whole studied area from the {\it Wide--field Infrared Survey Explorer (WISE)} with a 
simple color cut suggested by \citet{ste12}. Although the depth of {\it WISE} data is not enough to detect faint AGN, the wide area is helpful to measure the clustering 
after excluding bright AGN. We remove potential {\it WISE} AGN from our high sSFR galaxies, and then measure the clustering again. The fractions of galaxies excluded 
are 3.6\% in the LM bin and 8.9\% in the HM bin. The re-measured clusterings are identical to those for the previous measurements in both bins.
Since the triggering mechanism for AGN is not well understood yet, we cannot rule out the connection between high sSFR galaxies and AGN. However, 
our result for the highest sSFR bin is not biased by AGN.

Above, we noted that galaxies with the highest sSFR reside in more massive haloes than main sequence galaxies.
There are numerous reports in the literature that the star formation activity is enhanced by merger or the tidal interaction of galaxies 
\citep{mih92,bar00,lam03,ell08,hwa11,ide12,kam13,pat13}. \citet{dad10} and \citet{gen10} also reported that starburst galaxies driven by mergers form 
stars more efficiently than normal star-forming galaxies. Furthermore, it has been thought that these processes more easily occur in 
dense environments such as galaxy groups \citep{mci08,per09,pip14}. So when combined with the observation that these galaxies
are dustier and form stars efficiently, it suggests that high sSFR galaxies can be generated 
by mergers or interaction. Also, since massive galaxies usually reside in massive haloes, high sSFR HM galaxies experience the processes in 
massive haloes imprinting the $\sim$0.4 dex high halo mass compared to high sSFR LM galaxies as seen
in Figure~\ref{figmssfrmb}. This is consistent with \citet{lin12} who speculated that galaxies with the highest sSFR might be linked 
to mergers or interactions in haloes more massive than those hosting main sequence galaxies. Another possible explanation for high sSFR galaxies is 
that they are primarily in massive haloes like galaxy with enhanced star formation at $z\sim2$. \citet{mag14} pointed out that galaxies forming 
stars actively at $z\sim2$ are a different population from similar objects at $z<1$, based on the comparison of clustering strengths. They reported 
the halo mass of $>10^{13}M_{\odot}$, and they evolve into passive galaxies at $z<1.5$. The halo mass is consistent with that for our high sSFR samples. 
Moreover, \citet{pop15} also find that most IR-luminous galaxies at $z>1$ reside in the group environment, which is also similar to that for our high sSFR galaxies. 
In this sense, our high sSFR galaxies may be in the transition phase experiencing a rapid evolution to a passive population.

Now, we recall the HODs in Figure~\ref{figmbin} to interpret Figure~\ref{figmssfrmb}. First, we briefly note 
the halo mass difference measured by the halo model and the fit of dark matter clustering, since we directly estimate the bias with the 
dark matter clustering for sSFR binned samples. As we discuss in Appendix~\ref{app2}, the halo mass from the dark matter clustering fit 
is between representative halo masses for central and satellite galaxies from the halo model. If there are more satellites, this becomes closer to 
that for satellites. Furthermore, $M_{eff}$ of dark matter haloes from the halo model is more significantly influenced by satellites. In 
Figure~\ref{figmssfrmb}, the horizontal dashed lines are $M_{eff}$ values for all galaxies in each stellar mass bin. 

For LM galaxies, $M_{eff}$ for all galaxies from the halo model is $10^{12.66}h^{-1}M_{\odot}$ which is at the higher halo mass regime than the HOD for 
central galaxies in Figure~\ref{figmbin}. Therefore, if the halo mass is larger than this $M_{eff}$, we are able to conclude that satellite 
galaxies dominantly influence the clustering. Of LM samples, galaxies with $-10<\rm log\,sSFR/yr^{-1}<-9$ show halo masses of $\sim10^{13} h^{-1}M_{\odot}$. 
This means that these low sSFR LM galaxies mainly consist of satellites. In contrast, the halo mass for $-8.85<\rm log\,sSFR/yr^{-1}<-8.5$ galaxies 
corresponding to the upper part of the main sequence is below $M_{eff}$, which implies that the significant fraction of them is central galaxies. 

However, in the case of HM galaxies, it is more complex than LM galaxies. From Figure~\ref{figmbin}, it can be seen that the central HOD spans 
a wide halo mass range. Moreover, the $M_{eff}$ value for all HM galaxies is $10^{12.83}h^{-1}M_{\odot}$, and this is just above the peak of the central 
HOD. Therefore, we can reasonably expect a significant contribution from central galaxies to the clustering. From Figure~\ref{figmssfrmb}, the overall 
trend for HM star-forming galaxies is similar to that for LM star-forming ones. 
However, all halo masses do not exceed the range of the central HOD of 
the middle panel in Figure~\ref{figmbin}, and correspond to halo masses of $N_{c}(M_{halo})>N_{s}(M_{halo})$, except the highest sSFR bin. This 
also confirms that the important contribution of central galaxies to the clustering. What we also find is that the lowest halo mass appears 
in $-9.25<\rm log\,sSFR/yr^{-1}<-8.85$, which is the lower sSFR bin than that for LM galaxies, and corresponds to the upper envelope of the star-forming main sequence 
in Figure~\ref{figmssfr}.

Another striking feature is the halo mass of passive galaxies. We already discussed the dependence on stellar masses in the previous section. 
However, the halo mass of passive galaxies is also similar to $M_{eff}$ of all galaxies in each stellar mass bin. This implies that the influence of central passive galaxies is also 
significant as well as satellites, even in the LM bin. Moreover, the halo masses for both stellar mass bins are within the range of central HODs in Figure~\ref{figmbin}. In fact, the fractions of 
passive galaxies are 24\% and 50\% in the LM and HM bins, respectively. Even if we assume that all satellites are passive galaxies, 
these values are much larger than the satellite fraction for all galaxies from the HOD in Figure~\ref{figparbin}, and confirm the substantial contribution of central 
passive galaxies to the clustering. Additionally, \citet{kra13} found that red central galaxies tend to be in slightly 
more massive haloes than blue central galaxies based on SDSS data. Therefore, the higher halo mass for passive galaxies than main sequence ones can be 
explained by the high mass of haloes hosting passive central galaxies as well as passive satellites.

The complexity of Figure~\ref{figmssfrmb} highlights the potential of HOD analysis to identify subtle environmental and evolutionary trends if a sufficiently
large and well constrained sample of galaxies is studied. Based on our results for $M_{*}>10^{10}M_{\odot}$ galaxies at $z\sim1$, 
the clustering of main sequence galaxies, which shows the lowest halo mass, is most significantly affected by central galaxies. 
Given the hierarchical growth of structure and the expected evolution of main sequence star-forming galaxies, it would be expected that these central galaxies 
could become satellite galaxies via accretion on to massive haloes or remain as central galaxies with the assembly of stellar and halo masses. During this accretion, and 
whilst orbiting as satellites, they would undergo environmental 
quenching of their star formation (e.g., from ram pressure stripping, tidal harassment or starvation), which would explain the high halo masses associated with low sSFR 
galaxies.
On the other hand, the population with enhanced star formation activity are in relatively massive haloes of group like environments, 
and a subdominant population at $z\sim1$.
In our data, the fraction of $-8.5<\rm log\,sSFR/yr^{-1}<-8.0$ galaxies 
with $M_{*}>10^{10}M_{\odot}$ is $\sim$10\% of all galaxies, which is similar to the measured merger rates at $z=1$ from 
previous studies \citep{lot08,bun09,con09,der09,lop09}. Therefore, galaxy mergers play a subdominant role for the evolution of star-forming galaxies at this epoch, 
even if we assume all high sSFR galaxies experience mergers or interaction. 
Recently \citet{mou13} also concluded that galaxy mergers are not a dominant source of stellar mass growth at $z<1$.
Additionally, the significant fraction of passive galaxies are central ones which weakens the clustering strength compared to a satellite dominated sample.
These passive central galaxies may originate in star-forming central galaxies 
passively evolving, or galaxy mergers with the central galaxy accelerating the consumption of gas.

\citet{mos13} concluded that the secular processes are the dominant mechanism for the evolution of 
galaxies, which means that galaxies evolve from a low $M_{*}$--high sSFR to a high $M_{*}$--low sSFR regime
through star formation within the galaxy, and that galaxy mergers play a subdominant role. 
Moreover, \citet{tin13} claimed that the stellar mass function of passive central galaxies has 
significantly increased since $z=1$, especially at $M_{*}<10^{11}M_{\odot}$. Our results show that main sequence galaxies are 
mainly central galaxies, and a significant fraction of them evolve into low sSFR central galaxies together
with the decreased star formation activity and the increased stellar and halo masses.
Therefore, our result also supports the suggestion that the bulk of 
$M_{*}>10^{10}M_{\odot}$ star-forming galaxies at $z\sim1$ follow secular evolution (orange arrow in Figure~\ref{figmssfrmb}) 
supplemented by minor mergers with galaxies fainter than our stellar mass limit accounting for the bulk of their growth.
This was also suggested by \citet{noe07} and \citet{pen10}.
Additionally, a similar trend was found at $z\sim2$ \citep{lin12,sat14} and the local universe \citep{li08,hei09} which implies that this 
is the main evolutionary mechanism of star-forming galaxies over the last 10 Gyr. 
Importantly, the magnitude limit of our survey allows us to detect both galaxies that would give rise to a major 
merger within any halo so in principle it would be possible to constrain the major 
merger rate once the redshifts of a representative sub-sample of pairs is determined.

\section{SUMMARY AND CONCLUSION}\label{sumcon}

In this work, we have used deep and wide datasets based on UKIDSS DXS and CFHTLS--Wide surveys to 
investigate the dependence of galaxy clustering on intrinsic properties and how galaxies are linked with their 
host haloes. The main results are summarized as follows;

\begin{enumerate}

\item Using deep and wide near-IR/optical imaging data of the SA22 field, we have constructed a mass-limited 
sample of galaxies at $0.8<z<1.2$. The redshift selection is based on photometric redshifts, and galaxy 
properties such as stellar masses and sSFRs were derived using SED fitting. In total, this sample consists of 66,864 galaxies 
with $M_{*}>10^{10}M_{\odot}$ and $\rm log\,sSFR/yr^{-1}<-8$ in this redshift range.

\item Splitting the selected galaxies at different stellar mass thresholds, we measured the angular 
two-point correlation function and performed the halo modeling to link galaxies with host dark matter haloes. 
We found that more massive galaxies reside in more massive haloes, and tend to be central galaxies.

\item The HODs for stellar mass binned galaxies were calculated by those for stellar mass threshold 
samples. In all bins, satellites are predominantly in $>10^{13} h^{-1}M_{\odot}$ haloes. Moreover, the 
mass of haloes hosting central galaxies is higher for massive galaxies with a broader distribution 
in halo mass than that found for less massive galaxies.

\item The HODs for stellar mass binned samples were used to calculate the stellar mass to halo mass ratio 
for central and satellite galaxies separately. For central galaxies this ratio shows a peak at 
$\sim10^{12} h^{-1}M_{\odot}$ that drops sharply above and below this halo mass,  indicating the 
most efficient stellar mass growth at this peak halo mass. On the other hand, satellite galaxies significantly 
contribute to the total stellar mass in group and cluster environments.

\item We find an anti-correlation between bias and sSFR for low sSFR star-forming galaxies that are at or below 
the main sequence ($-10<\rm log\,sSFR/yr^{-1}<-8.5$), implying that star-forming galaxies at around the main 
sequence tend to live in a less massive halo ($M_{\rm halo}\sim10^{12.5}h^{-1}M_{\odot}$) while low sSFR or passive 
galaxies are more likely to be in massive haloes ($M_{\rm halo}>10^{12.5}h^{-1}M_{\odot}$). However, 
we also see a reversal of this relation for galaxies in the highest sSFR bin ($-8.5<\rm log\,sSFR/yr^{-1}<-8$) that 
galaxies with the highest sSFRs are in dense environments. This can be seen regardless of the 
stellar mass of galaxies.

\item We speculate that the bulk of galaxies evolves from on or above the main sequence of star-forming 
galaxies to a lower sSFR regime as their mass assembles through forming new stars and minor mergers. 
Additionally, major mergers happen in relatively massive haloes, and contribute to the galaxy evolution 
sub-dominantly at $z\sim1$.

\end{enumerate}

Our results are derived from deep and wide multiwavelength datasets. Nevertheless, this work is based on 
the only photometric data for relatively massive galaxies in a specific redshift bin and thus it is 
difficult to avoid contamination. In the future, deeper and wider datasets such as those from the UKIDSS, 
VISTA, Subaru Hyper-Suprime Camera, Pan-STARRS and LSST surveys will provide an opportunity to investigate more details 
about the relationship between various galaxy properties and their host dark matter haloes with less massive 
galaxies and in various redshift bins. Additionally, spectroscopic surveys will also allow further progress on the 
clustering analysis. In terms of parameterized HODs, we have assumed that the central HOD becomes unity after 
a certain halo mass. However, AGN feedback may change the shape of central HODs, especially the maximum mean 
number for central galaxies. Therefore it may also be necessary to modify the standard HOD work in order to 
directly compare observations and models once the samples have increased sufficiently.

\acknowledgments

Authors thank referee for valuable comments improving the presentation and content of the paper.
This work was supported by the National Research Foundation of Korea (NRF) grant, No. 2008-0060544, funded 
by the Korea government (MSIP). 
ACE acknowledges support from STFC grant ST/I001573/1.
We are grateful to UKIDSS team, the staff in UKIRT, Cambridge Astronomical Survey Unit and Wide Field
Astronomy Unit in Edinburgh. The United Kingdom Infrared Telescope is run by the Joint Astronomy Centre on 
behalf of the Science and Technology Facilities Council of the U.K. Based on observations obtained with 
MegaPrime/MegaCam, a joint project of CFHT and CEA/DAPNIA, at the Canada-France-Hawaii Telescope (CFHT) 
which is operated by the National Research Council (NRC) of Canada, the Institut National des Science 
de l'Univers of the Centre National de la Recherche Scientifique (CNRS) of France, and the University of 
Hawaii. This work is based in part on data products produced at the Canadian Astronomy Data Centre as part 
of the Canada-France-Hawaii Telescope Legacy Survey, a collaborative project of NRC and CNRS.



{\it Facilities:} \facility{UKIRT}, \facility{CFHT}

\begin{table*}
\begin{center}
\caption{The HOD parameters for stellar mass threshold galaxies at $0.8<z<1.2$. Column (1) represents the 
stellar mass threshold for each sub-sample in $M_{\odot}$, columns (2-5) are the best fit HOD parameters and 
column (6) is the number density of galaxies in $10^{-4} h^{3}$Mpc$^{-3}$. Columns (7-9) show the derived 
quantities base on equations (5)--(7). The final column is the quality of the HOD fit in terms of $\chi^{2}$ 
per degree of freedom. All dark matter halo masses are in $h^{-1}M_{\odot}$ with a logarithmic scale. 
\label{tparam}}
\resizebox{\linewidth}{!}{
\begin{tabular}{cccccccccc}
\tableline\tableline
Threshold & $\sigma_{cut}$ & $M_{cut}$ & $M_{0}$ & $\alpha$ & $n_{g}$ & $b_{g}$ & $M_{eff}$ & $f_{cen}$ & 
$\chi^{2}/dof$ \\
(1) & (2) & (3) & (4) & (5) & (6) & (7) & (8) & (9) & (10)\\
\tableline
$10^{10.0}$ & 0.50$^{+0.02}_{-0.07}$ & 11.885$^{+0.036}_{-0.049}$ & 12.912$^{+0.011}_{-0.004}$ & 1.14$^{+0.03}_{-0.02}$ & 68.8 & 1.62$^{+0.04}_{-0.04}$  & 12.778$^{+0.041}_{-0.040}$ & 0.84$^{+0.01}_{-0.01}$ & 3.47 \\
$10^{10.5}$ & 0.40$^{+0.03}_{-0.07}$ & 12.163$^{+0.046}_{-0.064}$ & 13.215$^{+0.015}_{-0.007}$ & 1.20$^{+0.04}_{-0.04}$ & 30.1 & 1.83$^{+0.07}_{-0.05}$  & 12.892$^{+0.053}_{-0.042}$ & 0.87$^{+0.01}_{-0.02}$ & 2.62 \\
$10^{11.0}$ & 0.60$^{+0.04}_{-0.04}$ & 12.958$^{+0.060}_{-0.049}$ & 13.909$^{+0.039}_{-0.036}$ & 0.90$^{+0.05}_{-0.06}$ &  4.7 & 2.25$^{+0.08}_{-0.08}$  & 13.117$^{+0.038}_{-0.038}$ & 0.94$^{+0.02}_{-0.03}$ & 1.27 \\
\tableline
\end{tabular}
}
\end{center}
\end{table*}

\begin{table}[h]
\begin{center}
\caption{Results for sSFR binned galaxies with stellar mass thresholds. Column (1) is the stellar mass threshold 
in $M_{\odot}$, column (2) shows the sSFR range with a logarithmic scale in $yr^{-1}$ and column (3) is the 
number of galaxies in each bin. Column (4-5) are the measured bias and the estimated halo mass, respectively. 
The unit of halo masses is in $h^{-1}M_{\odot}$.\label{tssfr}}
\resizebox{\columnwidth}{!}{
\begin{tabular}{ccccc}
\tableline\tableline
$M_{*}$ & sSFR range & $N_{gal}$ & bias & $M_{\rm halo}$ \\
(1) & (2) & (3) & (4) & (5) \\
\tableline
$M_{*}>10^{10}$ & -8.50 -- -8.00  & 6387  & 2.01$^{+0.08}_{-0.07}$ & 12.785$^{+0.064}_{-0.057}$ \\
                & -8.85 -- -8.50  & 9219  & 1.89$^{+0.07}_{-0.07}$ & 12.684$^{+0.064}_{-0.061}$ \\
                & -9.25 -- -8.85  & 9089  & 2.39$^{+0.09}_{-0.09}$ & 13.048$^{+0.056}_{-0.055}$ \\
                & -9.70 -- -9.25  & 6526  & 2.38$^{+0.11}_{-0.11}$ & 13.041$^{+0.069}_{-0.067}$ \\
                & -10.1 -- -9.70  & 4101  & 2.51$^{+0.11}_{-0.12}$ & 13.121$^{+0.064}_{-0.062}$ \\
                & $<$-10.5      & 24958 & 2.05$^{+0.05}_{-0.05}$ & 12.817$^{+0.038}_{-0.037}$ \\
$M_{*}>10^{10.5}$ & $<$-10.5      & 15863 & 2.22$^{+0.06}_{-0.06}$ & 12.940$^{+0.039}_{-0.038}$ \\
$M_{*}>10^{11.0}$ & $<$-10.5      & 3583  & 2.68$^{+0.10}_{-0.11}$ &  13.207$^{+0.053}_{-0.052}$\\
\tableline
\end{tabular}
}
\end{center}
\end{table}


\begin{table}
\begin{center}
\caption{The same table with Table~\ref{tssfr}, but for stellar mass binned galaxies. Column (1) represents 
the stellar mass range.\label{tssfrbi}}
\resizebox{\columnwidth}{!}{
\begin{tabular}{ccccc}
\tableline\tableline
$M_{*}$ & sSFR range & $N_{gal}$ & bias & $M_{\rm halo}$ \\
(1) & (2) & (3) & (4) & (5) \\
\tableline
$10^{10}<M_{*}<10^{10.5}$ & -8.50 -- -8.00  & 5082  & 1.90$^{+0.09}_{-0.09}$ & 12.694$^{+0.081}_{-0.077}$ \\
 (LM)                     & -8.85 -- -8.50  & 6675 & 1.79$^{+0.08}_{-0.09}$ & 12.585$^{+0.089}_{-0.083}$ \\
                          & -9.25 -- -8.85  & 6100 & 2.56$^{+0.10}_{-0.11}$ & 13.149$^{+0.059}_{-0.057}$ \\
                          & -9.70 -- -9.25  & 4049  & 2.50$^{+0.14}_{-0.15}$ & 13.114$^{+0.082}_{-0.079}$ \\
                          & -10.1 -- -9.70  & 2394  & 2.93$^{+0.17}_{-0.18}$ & 13.329$^{+0.079}_{-0.075}$ \\
                          & $<$-10.5        & 9095  & 1.85$^{+0.07}_{-0.08}$ & 12.643$^{+0.070}_{-0.067}$ \\
$10^{10.5}<M_{*}<10^{11}$ & -8.50 -- -8.00  & 1257  & 2.44$^{+0.18}_{-0.20}$ & 13.081$^{+0.115}_{-0.109}$ \\
  (HM)                    & -8.85 -- -8.50  & 2418 & 2.13$^{+0.14}_{-0.15}$ & 12.877$^{+0.110}_{-0.103}$ \\
                          & -9.25 -- -8.85  & 2807 & 1.94$^{+0.13}_{-0.14}$ & 12.725$^{+0.119}_{-0.110}$ \\
                          & -9.70 -- -9.25  & 2260  & 2.04$^{+0.16}_{-0.18}$ & 12.806$^{+0.138}_{-0.127}$ \\
                          & -10.1 -- -9.70  & 1527  & 2.25$^{+0.19}_{-0.21}$ & 12.961$^{+0.137}_{-0.127}$ \\
                          & $<$-10.5       & 12280 & 2.10$^{+0.06}_{-0.06}$ & 12.857$^{+0.046}_{-0.045}$ \\
\tableline
\end{tabular}
}
\end{center}
\end{table}


\appendix

\section{HOD MODELING WITHOUT $n_{g}$ CONSTRAINT}\label{app1}

As desribed in \S~\ref{hod}, the halo model with three free parameters was 
applied to derive the property of haloes hosting our galaxy sample. In this 
case, $M_{cut}$ was determined by matching the observed galaxy number density 
with given parameters. In order to check the influence of this constraint to 
the best fit result, we investigate the best fit parameters without this 
constraint. Therefore, we perform the halo modeling for 
$M_{*}>10^{10} M_{\odot}$ and $M_{*}>10^{11} M_{\odot}$ galaxies with 
four free parameters ($\sigma_{cut}$, $M_{cut}$, $M_{0}$ and $\alpha$). 
In this case, $n_{g}$ is derived by the best fit HOD parameters describing the 
clustering only.

\begin{figure}[h]
\centering
\includegraphics[scale=0.55]{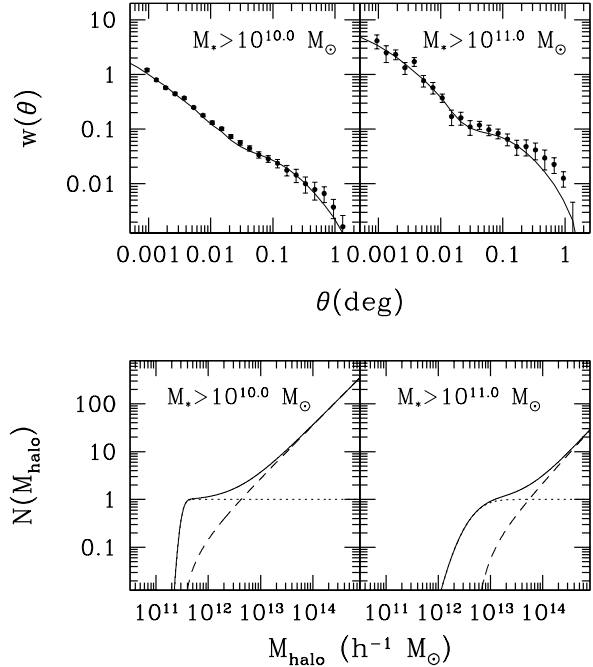}
\caption{The same plot with Figure~\ref{figmth} for $M_{*}>10^{10} M_{\odot}$ 
and $M_{*}>10^{11} M_{\odot}$ galaxies. All symbols are the same as 
Figure~\ref{figmth}. However, we used four free parameters for the halo 
modeling. As a result, the trend of derived parameters depending on $M_{*}$ is same to Figure~\ref{figmth}, 
although values are different.\label{fignong}}
\end{figure}

Figure~\ref{fignong} shows the angular correlation function (upper) and the 
HOD (lower) for each galaxy sample. All symbols are the same as 
Figure~\ref{figmth}. However, the solid line in the upper panel and the HOD in 
the lower panel are the result based on the halo model with four free parameters. 
The best fit parameters are listed in Table~\ref{tparanong}. Comparing the values 
in Table~\ref{tparam} and ~\ref{tparanong}, the derived values of $b_{g}$ and $M_{eff}$ are identical, and 
other halo mass parameters also show the same trend. In addition, the central HOD for massive 
galaxies ($\sigma_{cut}=$0.4) still shows the gentler shape than that for low mass galaxies ($\sigma_{cut}=$0.1), 
which indicates that 
the constraint by the galaxy number density does not affect the trend of the fitting results. 

However, $M_{cut}$ and $M_{0}$ without the $n_{g}$ constraint are approximately a factor of 2 smaller 
than those with the constraint, and $\sigma_{cut}$ values also decrease from 0.5 and 0.6 to 0.1 and 0.4 
for $M_{*}>10^{10} M_{\odot}$ and $M_{*}>10^{11} M_{\odot}$, respectively. In addition, the number densities 
calculated by the model without the constraint are higher than observed ones. This mismatch was already reported in 
\citet{mat11} and \citet{wak11}. Unfortunately, it is not yet answered what is the main reason leading to this 
mismatch. More studies are necessary to resolve this problem.

\clearpage

\section{HALO MODELING vs. FITTING DARK MATTER CLUSTERING}\label{app2}

In order to link galaxies to their host dark matter haloes, we performed the halo modeling for stellar mass threshold samples and the direct fit of correlation functions of dark matters for 
stellar mass binned samples and sSFR binned samples. Here, we apply the later method to the stellar mass threshold samples, and then compare bias and halo mass 
from this fit (quoted as DM fit, hereafter) to the best fit result from the halo model. This comparison will provide a guideline for our analysis. 

\begin{figure}[h]
\centering
\includegraphics[scale=0.39]{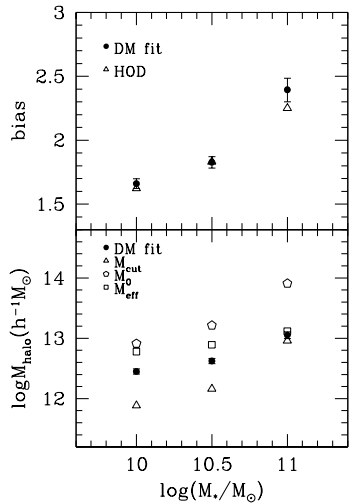}
\caption{Comparison of results from the halo model (open symbols) and the fit of dark matter correlation function (filled symbols) for 
stellar mass threshold samples. Upper and lower panels are for bias and halo mass, respectively. For the display purpose, 
the errors from the halo model are not displayed (see Table~\ref{tparam} for these values).
\label{figcomp2met}}
\end{figure}

Figure~\ref{figcomp2met} shows the comparison between 
the halo model (open symbols) in Table~\ref{tparam} and the fit of dark matter clustering (filled symbols). First, we are able to notice that the bias is consistent independently of the method (upper panel). 
However, in the case of halo masses, they show different results. The halo mass from the DM fit is close to $M_{eff}$ from the halo model, but shows a discrepancy, especially for low mass 
galaxies. This is easily explained by the fraction of satellites, since satellites are in massive haloes. In addition, $M_{eff}$ is close to $M_{0}$ at the low stellar mass regime, but 
$M_{cut}$ at the high mass regime, which is also explained by the same reason. 

Although the bias estimated by fitting the dark matter clustering do not perfectly represent the mass of haloes hosting galaxies selected, representative halo masses for central and satellite 
galaxies well bracket the halo mass by the DM fit. Therefore, the halo mass for central galaxies is always lower than that by the DM fit.

\begin{table*}
\begin{center}
\caption{The HOD parameters for stellar mass threshold galaxies. However, 
the halo modeling is performed with four free parameters without the constraint 
by the galaxy number density. All units are the same to table~\ref{tparam}.
\label{tparanong}}
\resizebox{\linewidth}{!}{
\begin{tabular}{cccccccccc}
\tableline\tableline
Threshold & $\sigma_{cut}$ & $M_{cut}$ & $M_{0}$ & $\alpha$ & $n_{g}$ & $b_{g}$ & $M_{eff}$ & $f_{cen}$ & 
$\chi^{2}/dof$ \\
(1) & (2) & (3) & (4) & (5) & (6) & (7) & (8) & (9) & (10)\\
\tableline
$10^{10.0}$ & 0.10$^{+0.0002}_{-0.0003}$ & 11.501$^{+0.001}_{-0.008}$ & 12.593$^{+0.012}_{-0.010}$ & 1.10$^{+0.02}_{-0.01}$ & 135.1$^{+5.8}_{-2.6}$ & 1.59$^{+0.03}_{-0.04}$  & 12.757$^{+0.036}_{-0.049}$ & 0.77$^{+0.02}_{-0.01}$ & 2.45 \\
$10^{11.0}$ & 0.40$^{+0.0277}_{-0.0278}$ & 12.692$^{+0.017}_{-0.021}$ & 13.697$^{+0.029}_{-0.027}$ & 1.20$^{+0.13}_{-0.10}$ &  7.1$^{+1.6}_{-1.1}$ & 2.26$^{+0.14}_{-0.14}$  & 13.115$^{+0.101}_{-0.088}$ & 0.92$^{+0.03}_{-0.03}$ & 1.25 \\
\tableline
\end{tabular}
}
\end{center}
\end{table*}

\end{document}